\documentclass[11pt]{article}

\usepackage[margin=1in]{geometry}
\usepackage{amsmath,amssymb}
\usepackage{booktabs}
\usepackage{graphicx}
\usepackage{caption}
\usepackage{subcaption}
\usepackage{float}
\usepackage{placeins}
\usepackage{authblk}
\usepackage{array}
\usepackage[authoryear,round]{natbib}
\usepackage{xcolor}
\usepackage{microtype}
\usepackage[ruled,vlined]{algorithm2e}
\usepackage{tikz}
\usetikzlibrary{arrows.meta,positioning}
\usepackage{hyperref}
\usepackage[utf8]{inputenc}

\hypersetup{colorlinks=true,linkcolor=blue,citecolor=blue,urlcolor=blue}
\usepackage{needspace}
\makeatletter
\renewcommand{\paragraph}{\@startsection{paragraph}{4}{\z@}{1.5ex plus 0.5ex minus 0.2ex}{0.7ex plus 0.2ex}{\normalfont\normalsize\bfseries}}
\makeatother

\title{\textbf{Simulation-Efficient Double-Flow Neural Posterior Estimation for Agent-Based Models}}
\author[1]{Junyao Zhang}
\author[1]{Jinglai Li}
\author[1]{Junqi Tang}
\affil[1]{School of Mathematics, University of Birmingham, Birmingham, United Kingdom}
\date{}

\begin{document}
\maketitle

\begin{abstract}
Neural posterior estimation (NPE) enables Bayesian inference of the posterior parameter in agent-based models (ABMs) with intractable likelihoods, but simulating training data can be computationally expensive. We propose Double-Flow Neural Posterior Estimation (DF-NPE), which combines learned time-series summaries with two conditional normalizing flows for synthetic-data generation and posterior estimation. The framework supplements direct simulations with synthetic training pairs and supports both pooled and ordered training strategies. Experiments in two settings of the Brock-Hommes asset-pricing model show that DF-NPE can reduce direct simulations by 90 percent while achieving lower mean posterior discrepancies and runtime than standard NPE. Performance varies with the allocation of directly simulated and synthetic data and the training strategy. These results establish learned summary augmentation as a practical approach to reducing simulation costs for Bayesian posterior inference of parameters in ABMs.
\end{abstract}

\noindent\textbf{Keywords:} Agent-based models; Bayesian inference; Simulation-based inference; Neural posterior estimation; Normalising flows; Synthetic data; Latent-surrogate augmentation

\section{Introduction}

Agent-based models (ABMs) are able to describe macroscopic henomena through the heterogeneous decisions and their interactions of individual agents. Applying these models in realistic analysis requires inference about their behavioural parameters \citep{WindrumEtAl2007,GrazziniEtAl2017}. Bayesian inference describes plausible parameter values and uncertainty about them through a posterior distribution. For many ABMs, however, the likelihood is intractable or expensive to evaluate. Simulation-based inference (SBI) uses model-simulated data to estimate the posterior, which requires the number and cost of simulator calls a practical constraint on calibration \citep{BeaumontEtAl2009,Lux2018}.

With the rapid development of machine learning, the neural posterior estimation (NPE) has been widely used in posterior inference in ABMs, which learns a conditional density from simulated parameter-data pairs \citep{CranmerEtAl2020}. Its sequential variant directs later simulations towards parameter regions relevant to a particular observation \citep{GreenbergEtAl2019}. Once fitted, the estimator generates posterior draws without further simulator calls. For ABMs, \citet{Shiono2021} applied BayesFlow to Bayesian parameter estimation, and \citet{DyerEtAl2024} studied neural inference with learned time-series summaries and evaluated recovery of the full posterior distribution. The remaining simulation cost lies in producing the training pairs, whose number must be weighed against the quality of the resulting posterior.

Nowadays, there has been more and more studies in SBI investigating approaches to reduce the cost of obtaining these simulated pairs or the need for them. Sequential proposals concentrate simulator calls in relevant parameter regions. When simulators at several fidelity levels are available, multilevel neural inference can improve accuracy under a fixed computational budget \citep{HikidaEtAl2025}. Surrogates approximate parts of the simulation or inference procedure: \citet{LampertiEtAl2018} use machine-learning surrogates to explore ABM parameter spaces, while \citet{Barde2024} uses Gaussian-process surrogates for Bayesian estimation. Neural emulators approximate simulator likelihoods \citep{LueckmannEtAl2019}, and sequential methods can jointly learn likelihood and posterior approximations \citep{WiqvistEtAl2021}. Additionally, for posterior training itself, \citet{RefaeliEtAl2025} use a neural density surrogate to replace simulator calls in later sequential rounds. Generated pairs can enlarge a training set, but they follow the learned surrogate distribution. Their contribution to posterior recovery therefore depends on the surrogate approximation and on how they are combined with direct simulations. We examine this question when the surrogate generates learned summaries and the representation changes across rounds.

In this paper, we investigate our Double-Flow neural posterior estimation (DF-NPE) framework using a recurrent encoder to map simulated trajectories to parameter-predictive summaries. The first conditional flow generates summaries given model parameters, and then the second estimates the parameter posterior given a summary. Generation takes place in this learned space rather than the full trajectory space. Similar to the sequential variants of NPE, at each sequential round, direct simulations update the encoder and surrogate. We then re-encode stored trajectories and regenerate synthetic summaries so that both data sources use the current representation. Within this procedure, we vary the allocation between directly-simulated and synthetic pairs and their order during posterior training. Mixed training pools the two sources, Synthetic-to-Simulation and Simulation-to-Synthetic use them in successive stages. Two settings of the Brock--Hommes model allow us to assess posterior recovery for heterogeneous forecasting parameters against approximate ground-truth posteriors obtained from tractable transition likelihoods. These comparisons extend the simulation-efficiency question studied by \citet{DyerEtAl2024} to the use of generated training data in sequential neural posterior estimation.

The contributions of this article are to provide:
\begin{enumerate}
\item a sequential DF-NPE procedure that combines learned time-series summaries, conditional latent generation and posterior estimation;
\item an evaluation of directly-simulated and synthetic data allocations through full-posterior discrepancies, predictive checks and recorded computational cost;
\item a comparison of posterior-training strategies and frozen-component controls that characterise when augmentation improves posterior recovery.
\end{enumerate}

\section{Double-Flow Neural Posterior Estimation}
\label{sec:framework}\label{sec:npe}\label{app:background}

Let $p(x\mid\theta)$ denote stochastic time-series data simulated from an ABM, $p(\theta)$ a prior distribution and $y_o$ an observed trajectory. Our aim is to estimate the posterior under a limited direct simulation budget,
\begin{equation}
p(\theta\mid y_o)\propto p(\theta)p(y_o\mid\theta).
\label{eq:bayes_target}
\end{equation}
NPE learns this conditional relationship from simulated parameter-data pairs, using summary statistics $s(x)$ to represent time-series observations \citep{PapamakariosMurray2016}. For prior draws $\theta_i\sim p(\theta)$ and trajectories $x_i\sim p(x\mid\theta_i)$, its training objective is
\begin{equation}
\widehat\psi=\arg\min_\psi -\frac{1}{N}\sum_{i=1}^{N}
\log q_\psi(\theta_i\mid s(x_i)).
\label{eq:npe_training}
\end{equation}
Conditioning the fitted estimator on $s(y_o)$ then yields posterior samples without further simulator calls. Sequential neural posterior estimation (SNPE) concentrates simulations in parameter regions relevant to the observation over successive rounds. Starting from $\pi_1=p(\theta)$, round $k$ generates pairs by sampling $\theta_i^{(k)}\sim\pi_k(\theta)$ and $x_i^{(k)}\sim p(x\mid\theta_i^{(k)})$. The fitted posterior then supplies the proposal for the next round. We use automatic posterior transformation, also known as SNPE-C, to account for the resulting non-prior proposals \citep{GreenbergEtAl2019}.

In our study, we propose a Double-Flow NPE (DF-NPE) framework which extends this procedure by learning a second conditional density that supplies additional training summaries. At round $k$, the recurrent encoder produces the learned summary statistics $h=e_{\omega_k}(x)\in\mathbb{R}^{8}$ and $h_{o,k}=e_{\omega_k}(y_o)$. We refer to $h$ as a latent summary, which the surrogate learns to generate. A latent-surrogate flow $q_{\phi_k}(h\mid\theta)$ generates summaries given parameters, while a posterior flow $q_{\psi_k}(\theta\mid h)$ learns the reverse conditional relationship, which are both based on masked autoregressive flows (MAFs) \citep{PapamakariosEtAl2017}. Normalising flows transform a simple base distribution through invertible mappings and evaluate the resulting density \citep{RezendeMohamed2015,cai2024nf}. Specifically, let $v=g_\xi(u)$, where $u\sim p_0$ and $f_\xi=g_\xi^{-1}$, according to the change-of-variables formula, we can obtain that
\begin{equation}
q_\xi(v)=p_0(f_\xi(v))\left|\det J_{f_\xi}(v)\right|,
\label{eq:flow_density}
\end{equation}
where the Jacobian determinant accounts for the local volume change induced by the transformation.

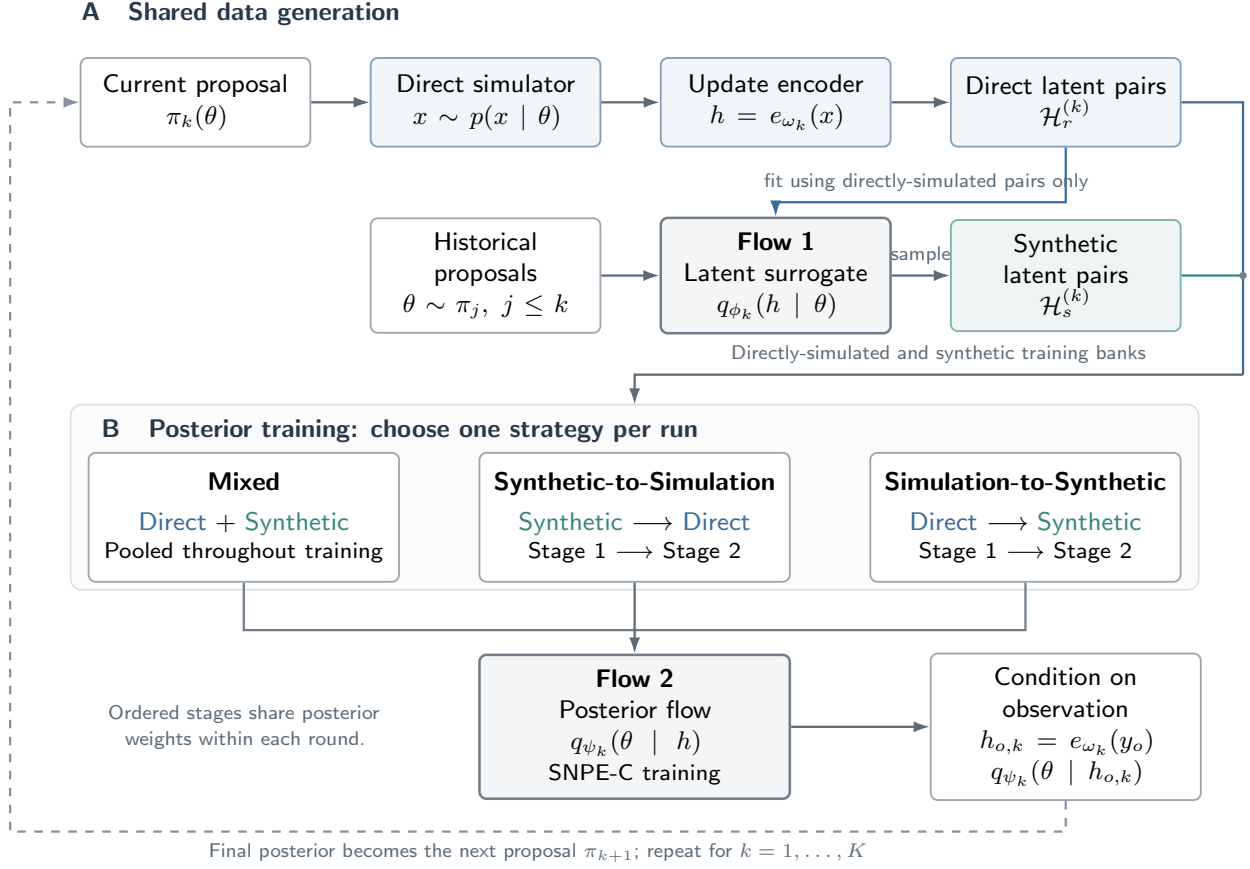
\begin{figure}[!htbp]
\centering
\resizebox{\textwidth}{!}{\begin{tikzpicture}[
  font=\sffamily\fontsize{9}{11}\selectfont,
  box/.style={draw=ink!45,line width=0.65pt,rounded corners=2.5pt,
    align=center,minimum height=11mm,text width=25mm,inner sep=5pt,fill=white},
  direct/.style={box,draw=directblue!65,fill=directblue!7},
  synthetic/.style={box,draw=syntheticteal!65,fill=syntheticteal!8},
  flow/.style={box,draw=ink!65,line width=0.9pt,fill=ink!5},
  strategy/.style={box,text width=35mm,minimum height=16mm,fill=white},
  arrow/.style={-{Latex[length=2mm,width=1.4mm]},draw=ink!75,line width=0.8pt},
  feedback/.style={arrow,dashed,draw=ink!55},
  note/.style={font=\sffamily\fontsize{7.5}{9}\selectfont,text=ink!75,align=center},
  paneltitle/.style={anchor=west,font=\sffamily\bfseries\fontsize{9}{11}\selectfont,text=ink}
]
\definecolor{ink}{HTML}{26394B}
\definecolor{directblue}{HTML}{3B6D9A}
\definecolor{syntheticteal}{HTML}{37877E}
\definecolor{warm}{HTML}{B58A4A}

\node[paneltitle] at (-1.55,1.10) {A\quad Shared data generation};
\node[box] (proposal) at (0,0) {Current proposal\\$\pi_k(\theta)$};
\node[direct] (simulator) at (3.6,0) {Direct simulator\\$x\sim p(x\mid\theta)$};
\node[direct] (encoder) at (7.2,0) {Update encoder\\$h=e_{\omega_k}(x)$};
\node[direct] (real) at (10.8,0) {Direct latent pairs\\$\mathcal H_r^{(k)}$};
\draw[arrow] (proposal) -- (simulator);
\draw[arrow] (simulator) -- (encoder);
\draw[arrow] (encoder) -- (real);

\node[box] (history) at (3.6,-2.15) {Historical proposals\\$\theta\sim\pi_j,\ j\leq k$};
\node[flow] (surrogate) at (7.2,-2.15) {\textbf{Flow 1}\\Latent surrogate\\$q_{\phi_k}(h\mid\theta)$};
\node[synthetic] (synth) at (10.8,-2.15) {Synthetic latent pairs\\$\mathcal H_s^{(k)}$};
\draw[arrow] (history) -- (surrogate);
\draw[arrow] (surrogate) -- node[above,note] {sample} (synth);
\draw[arrow,draw=directblue] (real.south) -- (10.8,-1.24) --
  node[above,note,pos=0.48] {fit using directly-simulated pairs only} (7.2,-1.24) -- (surrogate.north);

\draw[rounded corners=4pt,draw=ink!16,fill=ink!2,line width=0.6pt]
  (-1.55,-3.75) rectangle (12.45,-6.05);
\node[paneltitle] at (-1.30,-4.08) {B\quad Posterior training: choose one strategy per run};
\node[strategy] (mixed) at (0.60,-5.15)
  {\textbf{Mixed}\\[3pt]
   {\color{directblue}Direct} $+$ {\color{syntheticteal}Synthetic}\\[-1pt]
   {\fontsize{8}{9.5}\selectfont Pooled throughout training}};
\node[strategy] (ds) at (10.30,-5.15)
  {\textbf{Simulation-to-Synthetic}\\[3pt]
   {\color{directblue}Direct} $\longrightarrow$ {\color{syntheticteal}Synthetic}\\[-1pt]
   {\fontsize{8}{9.5}\selectfont Stage 1 $\longrightarrow$ Stage 2}};
\node[strategy] (sd) at (5.45,-5.15)
  {\textbf{Synthetic-to-Simulation}\\[3pt]
   {\color{syntheticteal}Synthetic} $\longrightarrow$ {\color{directblue}Direct}\\[-1pt]
   {\fontsize{8}{9.5}\selectfont Stage 1 $\longrightarrow$ Stage 2}};

\draw[draw=directblue,line width=0.8pt] (real.east) -- (13.0,0) -- (13.0,-3.38);
\draw[draw=syntheticteal,line width=0.8pt] (synth.east) -- (13.0,-2.15);
\fill[ink!65] (13.0,-2.15) circle (1.3pt);
\draw[arrow] (13.0,-3.38) -- node[above,note] {Directly-simulated and synthetic training banks} (5.45,-3.38) -- (5.45,-3.75);

\draw[draw=ink!65,line width=0.8pt] (mixed.south) -- (0.6,-6.55) -- (10.30,-6.55) -- (ds.south);

\node[flow,text width=35mm,minimum height=12mm] (posterior) at (5.45,-7.75)
  {\textbf{Flow 2}\\Posterior flow\\$q_{\psi_k}(\theta\mid h)$\\{\fontsize{8}{9.5}\selectfont SNPE-C training}};
\node[box,text width=30mm,minimum height=12mm] (final) at (10.80,-7.75)
  {Condition on observation\\$h_{o,k}=e_{\omega_k}(y_o)$\\$q_{\psi_k}(\theta\mid h_{o,k})$};
\draw[arrow] (sd.south) -- (posterior.north);
\draw[arrow] (posterior) -- (final);
\node[note,text width=35mm] at (0.6,-7.75)
  {Ordered stages share posterior weights within each round.};
\draw[feedback] (final.south) -- (10.80,-9.05) --
  node[below,note,pos=0.50] {Final posterior becomes the next proposal $\pi_{k+1}$; repeat for $k=1,\ldots,K$}
  (-2.30,-9.05) -- (-2.30,0) -- (proposal.west);
\end{tikzpicture}
}
\caption{DF-NPE with three posterior-training strategies. (A) Direct simulations train the encoder and latent-surrogate flow; stored trajectories are re-encoded, and the current surrogate generates synthetic summaries from retained parameter proposals. (B) The posterior flow is trained using Mixed, Synthetic-to-Simulation or Simulation-to-Synthetic, with the encoder and surrogate held fixed. Blue and teal denote directly-simulated and synthetic data, respectively. The dashed arrow indicates that the fitted posterior, conditioned on the observed summary, supplies the proposal for the next round.}
\label{fig:workflow}
\end{figure}

Additionaly, we also propose three strategies using the same sequence of simulation, summary learning, latent generation and sequential proposal updates for our DF-NPE framework. The choice of strategy enters at posterior training, where it determines the data-source order and stopping rule. We call simulator-generated trajectories and their encoded summaries \emph{directly-simulated}, and surrogate-generated latent pairs \emph{synthetic}. We also present the resulting sequential procedure in Figure~\ref{fig:workflow}.

\subsection{Flow 1: Learning summaries and generating synthetic data}
\label{sec:latent_generation}

In our DF-NPE, the first normalising flows are able to generate statistical summaries which avoids modelling complete trajectories, but the encoder must retain information about the parameters. We learn summaries by parameter regression, following the semi-automatic construction of \citet{FearnheadPrangle2012} and its neural-network implementation in \citet{JiangEtAl2017}. The recurrent embedding network in this work learns summary statistics from the temporal structure of the trajectories. The resulting posterior inference depends on the parameter information retained in these summaries \citep{BlumEtAl2013}. Although we can obtain summarizing statistic cheaper than directly calling the model simulator in this approach, more summary information does not automatically improve inference. For example, previous study in a Schelling segregation ABM showed that adding spatial or temporal summaries alone did not consistently help \citep{DignumEtAl2025}.

Let $\pi_j$ be the parameter proposal used in round $j$, with $\pi_1=p(\theta)$. The accumulated directly-simulated data comprise
\begin{equation}
\mathcal{D}_{r}^{(k)}=\bigcup_{j=1}^{k}
\{(\theta_{ji},x_{ji},j):\theta_{ji}\sim\pi_j,\quad
x_{ji}\sim p(\cdot\mid\theta_{ji})\}.
\end{equation}
The round index identifies the proposal used to generate each data block. A gated recurrent unit (GRU) encoder \citep{ChoEtAl2014} and an auxiliary head predict standardised parameters. Architecture and optimisation settings are given in Appendix~\ref{app:implementation}. At round $k$, the encoder and head minimise
\begin{equation}
\mathcal{L}_{\mathrm{enc}}(\omega,\eta)
=\frac{1}{n_r^{(k)}}\sum_{(\theta,x,j)\in\mathcal{D}_r^{(k)}}
\left\|a_\eta(e_\omega(x))-z_k(\theta)\right\|_2^2,
\end{equation}
where $n_r^{(k)}=|\mathcal{D}_r^{(k)}|$ and $z_k$ standardises parameters using the accumulated directly-simulated training set. After each encoder update, every stored trajectory is re-encoded. The resulting bank
\begin{equation}
\mathcal{H}_r^{(k)}=\{(\theta,e_{\omega_k}(x),j):
(\theta,x,j)\in\mathcal{D}_r^{(k)}\}
\end{equation}
expresses all directly-simulated summaries in the current representation. The observed trajectory is encoded with the same encoder. This shared representation provides the training space for the latent surrogate and the posterior flow.

We then fit the latent-surrogate MAF to the re-encoded directly-simulated bank by conditional maximum likelihood:
\begin{equation}
\mathcal{L}_{\mathrm{sur}}(\phi)
=-\frac{1}{n_r^{(k)}}\sum_{(\theta,h,j)\in\mathcal{H}_r^{(k)}}
\log q_\phi(h\mid\theta).
\end{equation}
The latent surrogate depends on the current encoder and on the parameter regions covered by direct simulations. Whenever either fitted component changes, we regenerate the synthetic summaries to keep them in the current representation. For each retained proposal $\pi_j$, synthetic pairs are generated as $\widetilde{\theta}_{ji}\sim\pi_j,\quad \widetilde{h}_{ji}\sim q_{\phi_k}(\cdot\mid\widetilde{\theta}_{ji})$. Their union forms $\mathcal{H}_s^{(k)}$, with each block retaining its proposal identity. We use these synthetic pairs only to train the posterior flow. The encoder and latent surrogate are fitted exclusively to direct simulator output.

\subsection{Flow 2: Posterior learning with directly-simulated and synthetic data}
\label{sec:schedules}

In our DF-NPE, we obtain the directly-simulated and synthetic summary banks after the first flow, which supply the training pairs for the posterior MAF. Training uses sequential neural posterior estimation with automatic posterior transformation \citep{GreenbergEtAl2019}, implemented with the \texttt{sbi} toolkit \citep{TejeroCanteroEtAl2020}. In the prior round, a minibatch $\mathcal B$ has loss
\begin{equation}
\mathcal L_{\mathrm{prior}}(\psi;\mathcal B)
=-\frac{1}{|\mathcal B|}\sum_{i\in\mathcal B}\log q_\psi(\theta_i\mid h_i).
\label{eq:posterior_prior_loss}
\end{equation}
Later rounds use the atomic loss. For each $(\theta_i,h_i)$, the candidate set $\mathcal A_i$ contains the paired parameter $\theta_i$ and contrasting parameters drawn from the minibatch. Defining the log score $r_\psi(\theta,h)=\log q_\psi(\theta\mid h)-\log p(\theta)$ gives
\begin{equation}
\mathcal L_{\mathrm{atom}}(\psi;\mathcal B)
=-\frac{1}{|\mathcal B|}\sum_{i\in\mathcal B}
\left[r_\psi(\theta_i,h_i)
-\log\sum_{\vartheta\in\mathcal A_i}\exp r_\psi(\vartheta,h_i)\right].
\label{eq:posterior_atomic_loss}
\end{equation}
The prior-density adjustment distinguishes this objective from uncorrected conditional likelihood on non-prior simulations. Each data block retains its generating proposal when the training source changes. This correction addresses parameter sampling; representation loss and surrogate error require the separate diagnostics reported below.

Directly-simulated pairs anchor training to simulator output, while synthetic pairs add draws from the learned conditional summary distribution. Their allocation determines bank composition; their training schedule determines how the posterior estimator encounters each source. Appendix~\ref{app:implementation} characterises the resulting augmented distribution. Let $\widehat P_{r,k}$ and $\widehat P_{s,k}$ denote the empirical distributions over directly-simulated and synthetic pairs, including their proposal provenance. At posterior update $t$, a source weight $\lambda_{k,t}\in[0,1]$ specifies
\begin{equation}
\widehat P_{k,t}=\lambda_{k,t}\widehat P_{r,k}
+(1-\lambda_{k,t})\widehat P_{s,k}.
\label{eq:sampling_schedule}
\end{equation}
This equation specifies source sampling rather than a decomposition of the atomic loss: pooled minibatches can contain contrastive parameters from both sources. With $n_s^{(k)}=|\mathcal H_s^{(k)}|$, the direct fraction is
\begin{equation}
\rho_k=\frac{n_r^{(k)}}{n_r^{(k)}+n_s^{(k)}}.
\label{eq:direct_fraction}
\end{equation}
Mixed training shuffles the pooled bank throughout optimisation, so
\begin{equation}
\lambda_{k,t}=\rho_k,\qquad t=1,\ldots,U,
\label{eq:mixed_schedule}
\end{equation}
where $U$ is determined by validation-based early stopping. Individual minibatches need not reproduce the exact global source ratio.

For ordered training, let $U_r$ and $U_s$ denote the directly-simulated and synthetic update budgets. Synthetic-to-Simulation first uses generated pairs and then continues from the resulting weights using directly-simulated pairs:
\begin{equation}
\lambda_{k,t}=\begin{cases}
0,&1\le t\le U_s,\\
1,&U_s<t\le U_s+U_r.
\end{cases}
\label{eq:synthetic_to_simulation}
\end{equation}
Simulation-to-Synthetic reverses the stages while retaining their source-specific update budgets:
\begin{equation}
\lambda_{k,t}=\begin{cases}
1,&1\le t\le U_r,\\
0,&U_r<t\le U_r+U_s.
\end{cases}
\label{eq:simulation_to_synthetic}
\end{equation}
Both stages are repeated within every round, with $U_r=U_s=500$ in the ordered experiments. We also provide more implementation details in Appendix~\ref{app:schedules}.

\subsection{Sequential inference and simulation budgets}

The training strategies of DF-NPE operate within the same sequential loop. Each round updates the encoder, re-encodes stored trajectories and regenerates the synthetic bank. During posterior training, the encoder and surrogate remain fixed. In the ordered implementation, posterior weights continue between stages, but each stage resets Adam. A fresh posterior flow is initialised in the next round because the encoder changes. After posterior training, the final estimate determines the next proposal
\begin{equation}
\pi_{k+1}(\theta)=q_{\psi_k^{\mathrm{final}}}(\theta\mid h_{o,k}).
\label{eq:proposal_update}
\end{equation}
We retain each proposal together with its conditioning summary and normalisation. The training strategy consequently affects later proposals and training data as well as the current posterior. Comparisons between strategies therefore evaluate complete sequential procedures.

These repeated updates also determine the computational budget. With $K$ rounds and final bank sizes $(N_r,N_s)$, each round adds $N_r/K$ direct simulations. At round $k$, the bank sizes are
\begin{equation}
n_r^{(k)}=\frac{kN_r}{K},\qquad n_s^{(k)}=\frac{kN_s}{K}.
\label{eq:round_bank_sizes}
\end{equation}
These expressions assume that the budgets are divisible by $K$ and count only valid accepted pairs. The final bank has $N=N_r+N_s$ pairs and direct fraction $\rho=N_r/N$. Because synthetic banks are regenerated, the total accepted synthetic generation is
\begin{equation}
\sum_{k=1}^{K}n_s^{(k)}=\frac{K+1}{2}N_s.
\end{equation}
For ten rounds this equals $5.5N_s$, rather than $N_s$. Direct simulator calls, final bank sizes, regenerated pairs and optimisation updates therefore measure different aspects of computational cost. Here, each round also updates the encoder and latent surrogate before posterior training. The encoder is then held fixed, whereas the embedding network in their algorithm is learned jointly with the posterior estimator. The three branches in Algorithm~\ref{alg:latent_surrogate_snpe} specify the data source and stopping rule for each posterior-training stage. In the algorithm, $B_k=200$ for Mixed and $B_k=\min\{200,n_r^{(k)},n_s^{(k)}\}$ for ordered training. The stage lists are rebuilt in every round, and the proposal is updated only after all stages finish. More details are provided in Appendix~\ref{app:implementation}.

\begin{algorithm}[!htbp]
\caption{Sequential DF-NPE with three training strategies}
\label{alg:latent_surrogate_snpe}
\small
\DontPrintSemicolon
\KwIn{Prior $p(\theta)$, simulator $p(x\mid\theta)$, observation $y_o$, rounds $K$, final bank sizes $(N_r,N_s)$, strategy $\mathsf S$, minibatch size $B_k$ and ordered-stage budgets $U_r=U_s=500$.}
\KwOut{Posterior estimate $q_{\psi_K^{\mathrm{final}}}(\theta\mid h_{o,K})$.}
Set $\pi_1=p(\theta)$ and $\mathcal D_r^{(0)}=\varnothing$\;
\For{$k=1,\ldots,K$}{
    Draw $N_r/K$ pairs $\theta\sim\pi_k$, $x\sim p(\cdot\mid\theta)$;
    set $\mathcal D_r^{(k)}=\mathcal D_r^{(k-1)}\cup\{(\theta_i,x_i,k)\}_{i=1}^{N_r/K}$\;
    Update $e_{\omega_k}$ by parameter regression; re-encode all stored trajectories as $\mathcal H_r^{(k)}$ and set $h_{o,k}=e_{\omega_k}(y_o)$\;
    Fit $q_{\phi_k}(h\mid\theta)$ by conditional maximum likelihood on $\mathcal H_r^{(k)}$ only\;
    Rebuild $\mathcal H_s^{(k)}$: for each $j\le k$, draw $N_s/K$ pairs $\theta\sim\pi_j$, $h\sim q_{\phi_k}(\cdot\mid\theta)$; retain the source-proposal label $j$\;
    Initialise a posterior flow $q_\psi$ with pooled-bank normalisation; keep the encoder, surrogate and normalisation fixed during posterior training\;
    \uIf{$\mathsf S=\textnormal{Mixed}$}{
        Set the stage list $\mathcal T_k=[\mathcal H_r^{(k)}\cup\mathcal H_s^{(k)}]$; shuffle the pooled bank and use validation-based early stopping\;
    }
    \uElseIf{$\mathsf S=\textnormal{Synthetic-to-Simulation}$}{
        Set $\mathcal T_k=[\mathcal H_s^{(k)},\mathcal H_r^{(k)}]$; use $U_s$ synthetic updates followed by $U_r$ direct updates\;
    }
    \Else{
        Set $\mathcal T_k=[\mathcal H_r^{(k)},\mathcal H_s^{(k)}]$ for Simulation-to-Synthetic; use $U_r$ direct updates followed by $U_s$ synthetic updates\;
    }
    \ForEach{bank $\mathcal H\in\mathcal T_k$, in the listed order}{
        Attach retained proposals to the data blocks in $\mathcal H$; shuffle the bank and initialise Adam for this stage\;
        \Repeat{the selected stopping rule is met for this stage}{
            Draw a minibatch $\mathcal B=\{(\theta_i,h_i,j_i)\}$ of size $B_k$ from $\mathcal H$\;
            \eIf{$k=1$}{
                Compute the conditional negative log-likelihood:
                $\displaystyle \mathcal L=-|\mathcal B|^{-1}\sum_{i\in\mathcal B}\log q_\psi(\theta_i\mid h_i)$\;
            }{
                For each $i$, form $\mathcal A_i$ from $\theta_i$ and contrasting parameters sampled from the other minibatch entries\;
                Evaluate $r_{ia}=\log q_\psi(\theta_{ia}\mid h_i)-\log p(\theta_{ia})$ for $\theta_{ia}\in\mathcal A_i$, with $\theta_{i1}=\theta_i$\;
                Compute the atomic classification loss:
                $\displaystyle \mathcal L=-|\mathcal B|^{-1}\sum_{i\in\mathcal B}\left[r_{i1}-\log\sum_a\exp(r_{ia})\right]$\;
            }
            Take an Adam update of $\psi$ using $\nabla_\psi\mathcal L$\;
        }
        Carry $\psi$ into the next stage without reinitialising the posterior flow, if a second stage remains\;
    }
    Condition the final posterior on $h_{o,k}$ and set
    $\pi_{k+1}(\theta)=q_{\psi_k^{\mathrm{final}}}(\theta\mid h_{o,k})$;
    retain the proposal with its conditioning, prior support and normalisation\;
}
\end{algorithm}
\FloatBarrier

\section{Numerical Experiments}
\label{sec:experiments}\label{sec:results}\label{sec:mixed_experiments}

In this section, we evaluate DF-NPE in two settings of the Brock--Hommes model, following the benchmark studied by \citet{DyerEtAl2024}. Because the model has a tractable transition likelihood, we can obtain its approximate ground-truth posteriors allowing us to assess how well DF-NPE recovers parameter uncertainty. We first vary the proportion of directly-simulated and synthetic pairs to examine posterior recovery under different simulation budgets. We then compare the three posterior-training strategies at shared allocations and assess their computational cost. Posterior plots illustrate the recovered distributions, and boxplots show how discrepancies and runtime vary across repeated fits. Supplementary equal-budget comparisons, predictive checks and frozen-component controls also help interpret these results.

\subsection{Performance metrics}
\label{sec:metrics}

Assessing Bayesian calibration requires comparing posterior distributions as well as parameter estimates. We therefore use joint Wasserstein distance (WASS) and squared maximum mean discrepancy  (MMD$^2$) to evaluate posterior recovery. Both metrics compare the estimated posterior $Q$ with the approximate ground-truth posterior $P_{\mathrm{ref}}$ as reference for the same observed trajectory $y_o$. We compute them using reference draws $\{\theta_i\}_{i=1}^{n}$ and estimated draws $\{\vartheta_j\}_{j=1}^{m}$, with $n=m=1{,}000$. The resulting scores are empirical comparisons and retain Monte Carlo error from both samples. Posterior-mean error and posterior-predictive checks provide additional assessments, which will be further discussed in Appendix~\ref{app:score_details}.

\paragraph{Wasserstein distance}
Wasserstein distance measures the minimum cost of transporting mass between the two empirical posterior distributions \citep{BerntonEtAl2019}. We use first-order Wasserstein distance with Euclidean transport cost,
\begin{equation}
\operatorname{WASS}(\widehat P_{\mathrm{ref}},\widehat Q)
=\min_{\gamma\in\Gamma_{n,m}}\sum_{i=1}^{n}\sum_{j=1}^{m}
\gamma_{ij}\|\theta_i-\vartheta_j\|_2,
\label{eq:evaluation_wass}
\end{equation}
where $\Gamma_{n,m}$ contains non-negative transport matrices with row sums $1/n$ and column sums $1/m$. The transport cost is computed in the joint parameter space, so it responds to differences in location, spread and dependence. This distinguishes it from distances based only on posterior means or averaged univariate marginals. A lower value indicates closer agreement with the reference posterior.

\paragraph{Squared maximum mean discrepancy}
MMD compares kernel mean embeddings of probability distributions in a reproducing kernel Hilbert space $\mathcal H$ \citep{GrettonEtAl2012}. Its population squared form is
\begin{equation}
\operatorname{MMD}^2(P_{\mathrm{ref}},Q)
=\left\|\mathbb E_{\theta\sim P_{\mathrm{ref}}}[k(\theta,\cdot)]
-\mathbb E_{\vartheta\sim Q}[k(\vartheta,\cdot)]\right\|_{\mathcal H}^2.
\label{eq:evaluation_mmd_population}
\end{equation}
We estimate squared MMD by excluding the within-sample diagonal terms,
\begin{equation}
\begin{aligned}
\widehat{\operatorname{MMD}}_u^2
={}&\frac{1}{n(n-1)}\sum_{i\ne i'}k(\theta_i,\theta_{i'})
+\frac{1}{m(m-1)}\sum_{j\ne j'}k(\vartheta_j,\vartheta_{j'})-\frac{2}{nm}\sum_{i=1}^{n}\sum_{j=1}^{m}k(\theta_i,\vartheta_j).
\end{aligned}
\label{eq:evaluation_mmd_estimator}
\end{equation}
For a fixed kernel and independent draws, this is the usual unbiased estimator of squared MMD. We use a Gaussian kernel with a reference-based scale,
\begin{equation}
 k(a,b)=\exp\!\left(-\frac{\|a-b\|_2^2}{c}\right),
 \qquad c=2\operatorname{median}_{1\le i,i'\le n}\|\theta_i-\theta_{i'}\|_2.
\label{eq:evaluation_mmd_kernel}
\end{equation}
Here $c$ is based on median Euclidean distance, rather than the squared-distance convention used in some Gaussian kernels. Under this common kernel, a lower MMD$^2$ indicates closer posterior agreement, although the finite-sample estimate can be slightly negative. We retain the squared statistic throughout and report results separately for each setting and diagnostic kernel because their scales differ.

\paragraph{Computational cost.}
We assess computational efficiency through direct simulator calls and recorded end-to-end runtime. The latter includes simulation, neural training, synthetic generation, posterior sampling, input/output and plotting, but excludes separate sampling for posterior displays. We calculate speed-up by dividing the mean SNPE runtime by the corresponding workflow mean. For the smooth setting, accuracy and runtime were measured in separate runs, and timing comparisons were not interleaved. The recorded speed-ups consequently describe the implemented workflows under these execution conditions.

\subsection{Brock--Hommes model}
\label{sec:bh_model}

The Brock--Hommes (BH) model describes an asset market in which agents choose among heterogeneous forecasting strategies \citep{BrockHommes1998}. Each strategy $h$ has a trend coefficient $g_h$ and bias $b_h$, and its performance $U_{h,t}$ determines the fraction of agents using it. Writing the price deviation from fundamental value as $x_t$, the dynamics are
\begin{equation}
\begin{aligned}
U_{h,t} &= (x_t-Rx_{t-1})(g_hx_{t-2}+b_h-Rx_{t-1}),\\
n_{h,t+1} &= \frac{\exp(\beta U_{h,t})}{\sum_{j=1}^{H}\exp(\beta U_{j,t})},\\
x_{t+1} &= \frac{1}{R}\left\{\sum_{h=1}^{H}n_{h,t+1}(g_hx_t+b_h)+\epsilon_{t+1}\right\},
\quad \epsilon_{t+1}\sim\mathcal N(0,\sigma^2).
\end{aligned}
\label{eq:bh_dynamics}
\end{equation}
Here $n_{h,t+1}$ is the fraction using strategy $h$, $R$ is the gross risk-free return and $\beta$ controls choice intensity. We use four strategies and infer $\theta=(g_2,b_2,g_3,b_3)$. These parameters describe the trend responses and forecasting biases of two strategies; the remaining coefficients are fixed at $R=1.01$, $g_1=b_1=b_4=0$ and $g_4=1.01$ \citep{BrockHommes1998,DyerEtAl2024}. The resulting posterior describes which combinations of these behavioural coefficients are compatible with the observed price dynamics.

We consider two settings. In the smooth setting, $\beta=120$ and the generating parameter is $(0.9,0.2,0.9,-0.2)$; the independent priors are $g_2,b_2,g_3\sim\mathcal U(0,1)$ and $b_3\sim\mathcal U(-1,0)$. In the noisy setting, $\beta=10$ and the generating parameter is $(-0.7,-0.4,0.5,0.3)$; the independent priors are $g_2,b_2\sim\mathcal U(-1,0)$ and $g_3,b_3\sim\mathcal U(0,1)$. Both settings use $\sigma=0.04$, so ``noisy'' names the regime rather than an increase in innovation variance. Each simulator allocates 100 time steps and discards three deterministic initial values. Each setting is a separate inference task with its own fixed simulated observation, prior and reference posterior.

The Gaussian innovation in the price equation gives the conditional transition density
\begin{equation}
p(x_{t+1}\mid x_{1:t},\theta)
=\mathcal N\!\left(x_{t+1};\frac{1}{R}\sum_{h=1}^{H}n_{h,t+1}(g_hx_t+b_h),\frac{\sigma^2}{R^2}\right).
\end{equation}
We use this transition density to construct a approximate ground-truth posterior by likelihood-based MCMC. The neural procedures are trained on simulated parameter--data pairs without evaluating the likelihood, so the approximate ground-truth posterior provides a reference for assessing posterior recovery.

\subsubsection{Experimental setup}
\label{sec:evaluation}

Throughout the experiments, SNPE denotes the baseline method, and DF-NPE denotes the proposed framework. Mixed, Synthetic-to-Simulation and Simulation-to-Synthetic denote its three posterior-training strategies; thus, Mixed 3:7 denotes DF-NPE with Mixed training at a directly-simulated:synthetic ratio of 3:7. The approximate ground-truth posterior is obtained by likelihood-based MCMC and is subsequently called the reference posterior. In the boxplots, Mixed and the ordered-strategy labels identify DF-NPE configurations, while the label ``Standard SNPE'' denotes SNPE.

For each BH setting, we evaluate each main configuration across ten training seeds using a fixed simulated observation and 1,000 likelihood-based MCMC reference draws. Further experimental details are provided in Appendix~\ref{app:models}. All primary comparisons use ten sequential rounds and a final posterior-training bank of 10,000 pairs. Within this bank, Mixed allocates directly-simulated and synthetic pairs in ratios of 1:9, 3:7, 5:5, 7:3 or 9:1. These allocations require 1,000--9,000 direct simulations, compared with 10,000 for SNPE. The procedures also differ in representation learning, that is, SNPE learns its recurrent embedding jointly with the posterior, whereas DF-NPE uses the supervised encoder in Section~\ref{sec:framework}. Their comparison therefore evaluates both the training-data allocation and the representation-learning procedure.

To compare the procedures at equal direct simulation budgets, we conduct a separate experiment in the smooth setting with five paired seeds. In this experiment, SNPE uses 1,000--9,000 directly-simulated pairs, matching the direct component of each Mixed bank. We compare the ordered strategies at ratios 1:9, 3:7 and 5:5, giving 120 ordered fits across two settings and ten seeds. Each strategy uses 500 updates per source per round, with batch size $\min\{200,n_r^{(k)},n_s^{(k)}\}$. This fixed update budget differs from the validation-based early stopping used by Mixed. Moreover, each fitted posterior supplies the next proposal, so a change in training strategy also changes the subsequent training banks.

We summarise the ten fits with boxplots of the discrepancies and runtime, using their medians, interquartile ranges and outliers to describe performance across seeds. Appendix~\ref{app:additional_results} reports the corresponding means and sample standard deviations, whose rankings can differ from those of the medians. We also report descriptive paired-seed comparisons, while allowing each procedure to retain its own learned representation and subsequent proposals.

\subsubsection{Results}

\begin{figure}[!htbp]
\centering
\begin{subfigure}[t]{0.32\textwidth}\centering
\includegraphics[width=\linewidth]{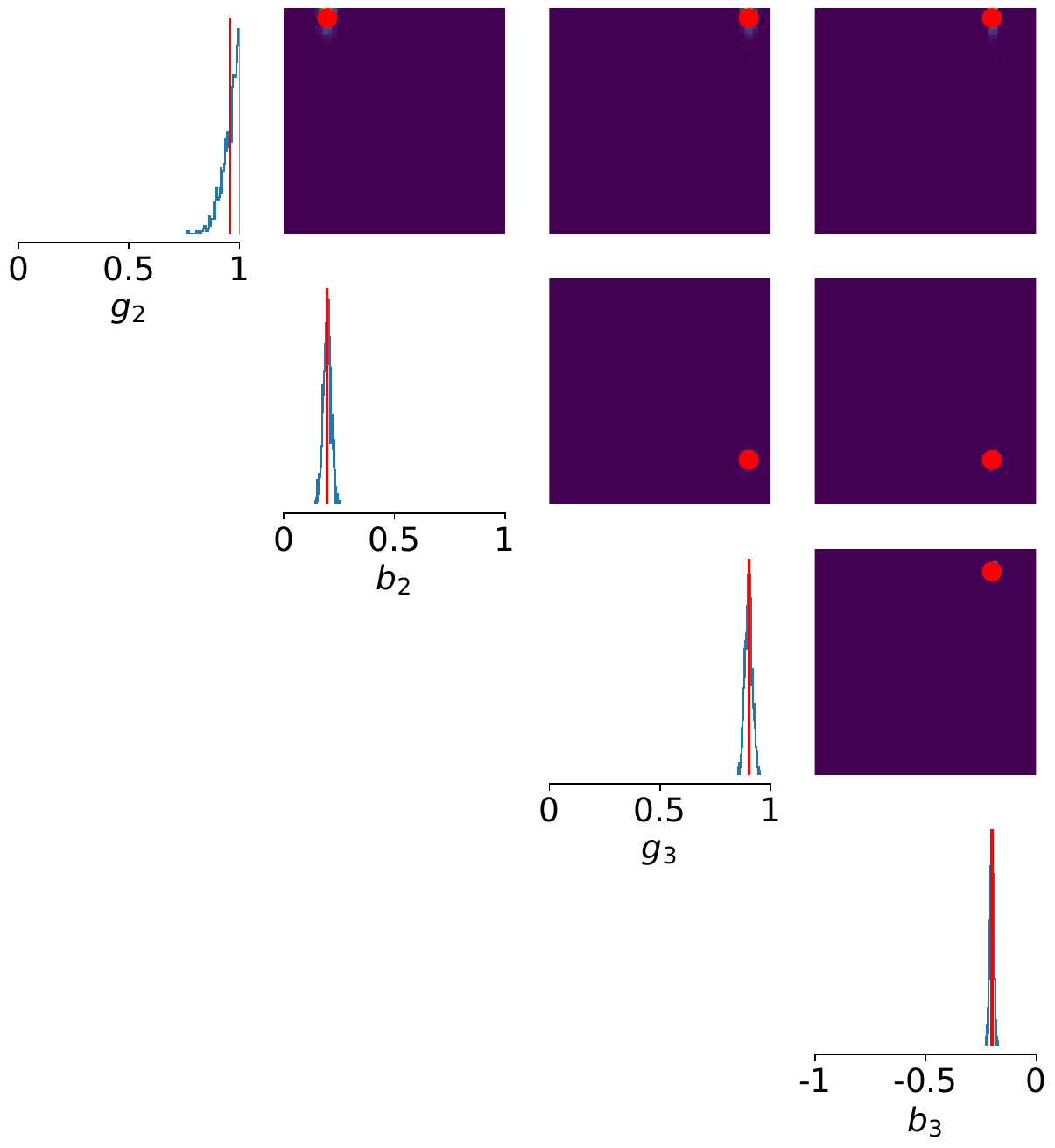}
\caption{Approximate ground truth}
\end{subfigure}
\hfill
\begin{subfigure}[t]{0.32\textwidth}\centering
\includegraphics[width=\linewidth]{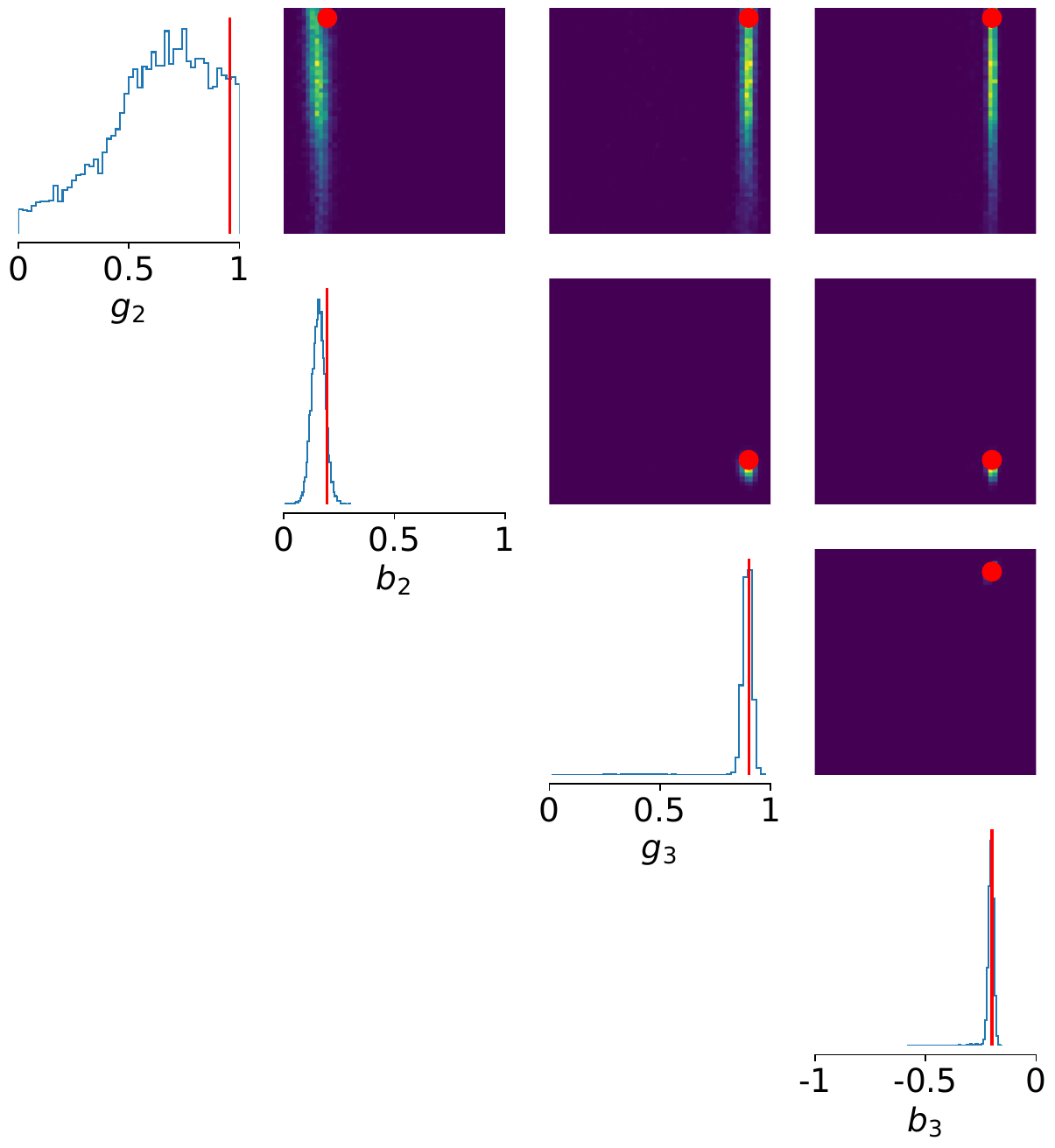}
\caption{SNPE}
\end{subfigure}
\hfill
\begin{subfigure}[t]{0.32\textwidth}\centering
\includegraphics[width=\linewidth]{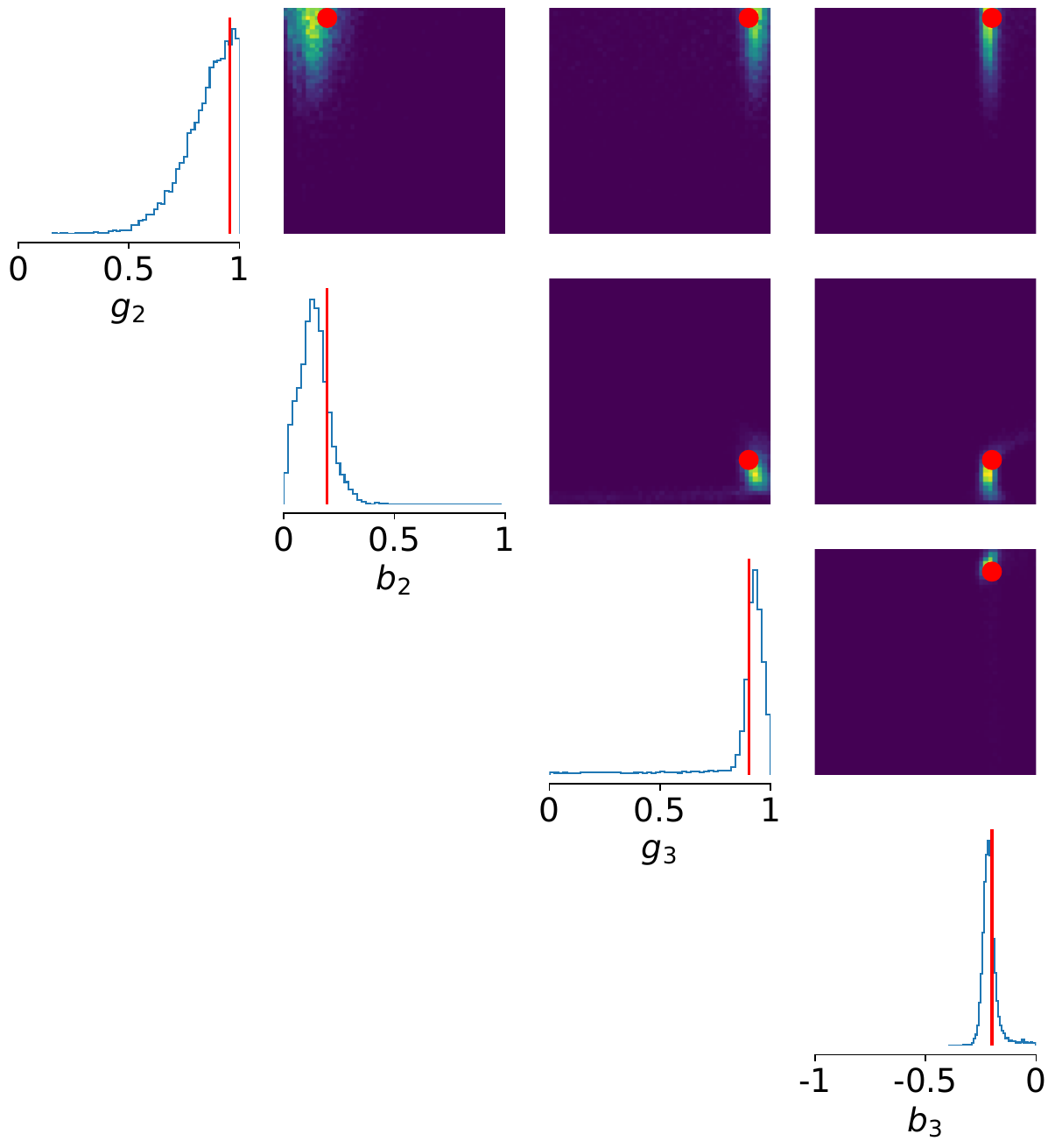}
\caption{DF-NPE}
\end{subfigure}
\caption{Posterior samples from the approximate ground truth, SNPE and DF-NPE for the smooth Brock--Hommes setting. DF-NPE uses Mixed training at a directly-simulated:synthetic ratio of 3:7. Diagonal panels show marginal posterior distributions as count histograms; upper-triangular panels show bivariate joint distributions as two-dimensional count histograms. Red lines and points denote the mean of the likelihood-based reference posterior. Darker and lighter colours indicate lower and higher sample counts, respectively, in the off-diagonal plots.}
\label{fig:bh_smooth_overview}
\end{figure}

\paragraph{Smooth setting.}\label{sec:bh_smooth}
In the smooth BH setting, the approximate ground-truth posterior concentrates $g_2$ and $g_3$ near their upper prior boundaries and has narrow marginals for $b_2$ and $b_3$. Figure~\ref{fig:bh_smooth_overview} compares this structure with SNPE and DF-NPE using Mixed training at a directly-simulated:synthetic ratio of 3:7. The most visible difference concerns $g_2$: the SNPE marginal extends towards smaller values, while Mixed places more mass near the upper boundary. Mixed nevertheless retains a broader $b_2$ marginal than the reference. These differences also appear in the bivariate distributions, where the concentration of posterior mass reflects the corresponding marginal shapes.

\begin{figure}[!htbp]
\centering
\begin{subfigure}[t]{0.96\textwidth}\centering
\includegraphics[width=\linewidth]{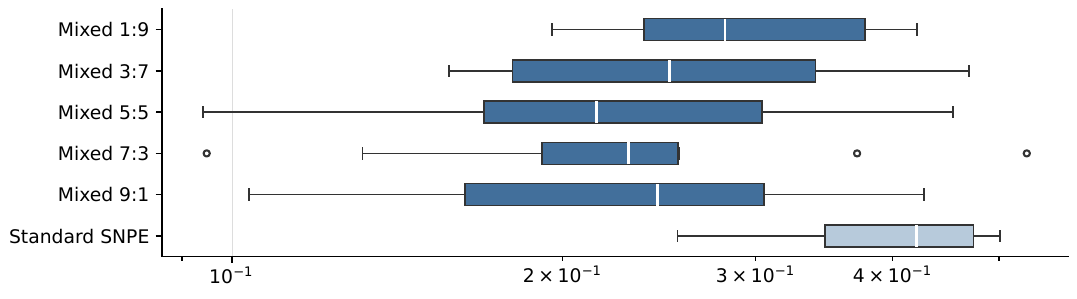}
\caption{Wasserstein distance.}\end{subfigure}\par\vspace{0.4em}
\begin{subfigure}[t]{0.96\textwidth}\centering
\includegraphics[width=\linewidth]{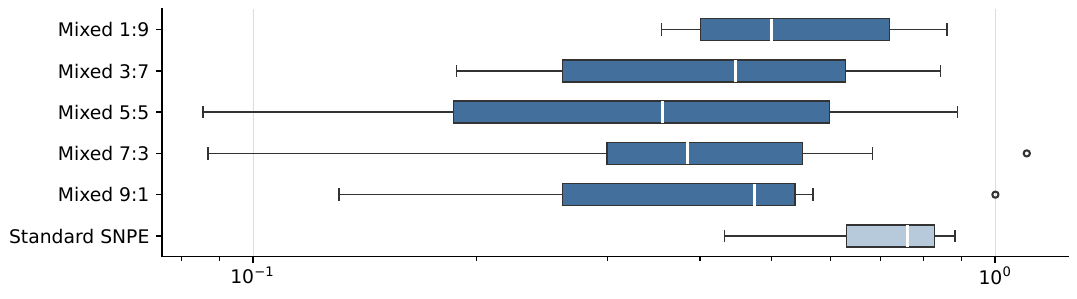}
\caption{Squared maximum mean discrepancy.}\end{subfigure}
\caption{Posterior discrepancies of SNPE and DF-NPE with Mixed training for the smooth Brock--Hommes setting. Results show WASS and MMD$^2$ over ten fits per configuration using one fixed observation. Ratios denote directly-simulated:synthetic allocations; SNPE uses only direct simulations. White lines denote medians, boxes show the interquartile range, and whiskers extend to the most extreme values within 1.5 interquartile ranges. }
\label{fig:mixed_fidelity}
\end{figure}

The repeated-fit results extend this comparison across all five allocations (Figure~\ref{fig:mixed_fidelity}). Every Mixed allocation had lower median WASS and MMD$^2$ than SNPE, including 1:9, which used 1,000 rather than 10,000 direct simulations. Increasing the direct fraction to 3:7 and then 5:5 lowered both medians further. However, the boxes overlapped and showed substantial variation across fits, so the improvement in median performance did not separate all allocations clearly.

Beyond 5:5, increasing the direct fraction produced no further reduction in either median. The 7:3 and 9:1 allocations also contained high-discrepancy outliers, indicating that a larger directly-simulated bank did not consistently improve recovery. The mean scores give a slightly different ranking: 9:1 had the lowest mean WASS, while 5:5 had the lowest mean MMD$^2$ (Table~\ref{tab:historical_all}). Thus, although all allocations improved on the baseline in both means and medians, their relative ranking depended on the metric and summary statistic.

The comparisons above hold the total bank size fixed while changing the number of direct simulations. When we instead matched the directly-simulated budget, Mixed reduced WASS in all 25 paired comparisons and MMD$^2$ in 23 (Appendix~\ref{app:matched_budget}). This supports the complete DF-NPE procedure under the second budget comparison, including its different approach to representation learning.

Posterior agreement does not by itself determine how closely simulated trajectories reproduce the observation. We therefore also performed posterior-predictive checks in the smooth setting (Appendix~\ref{app:predictive_scores}). SNPE had the lowest median-trajectory RMSE, whereas Mixed 5:5 had the lowest summary error, showing that these predictive criteria ranked the procedures differently.

\begin{figure}[!htbp]
\centering
\begin{subfigure}[t]{0.32\textwidth}\centering
\includegraphics[width=\linewidth]{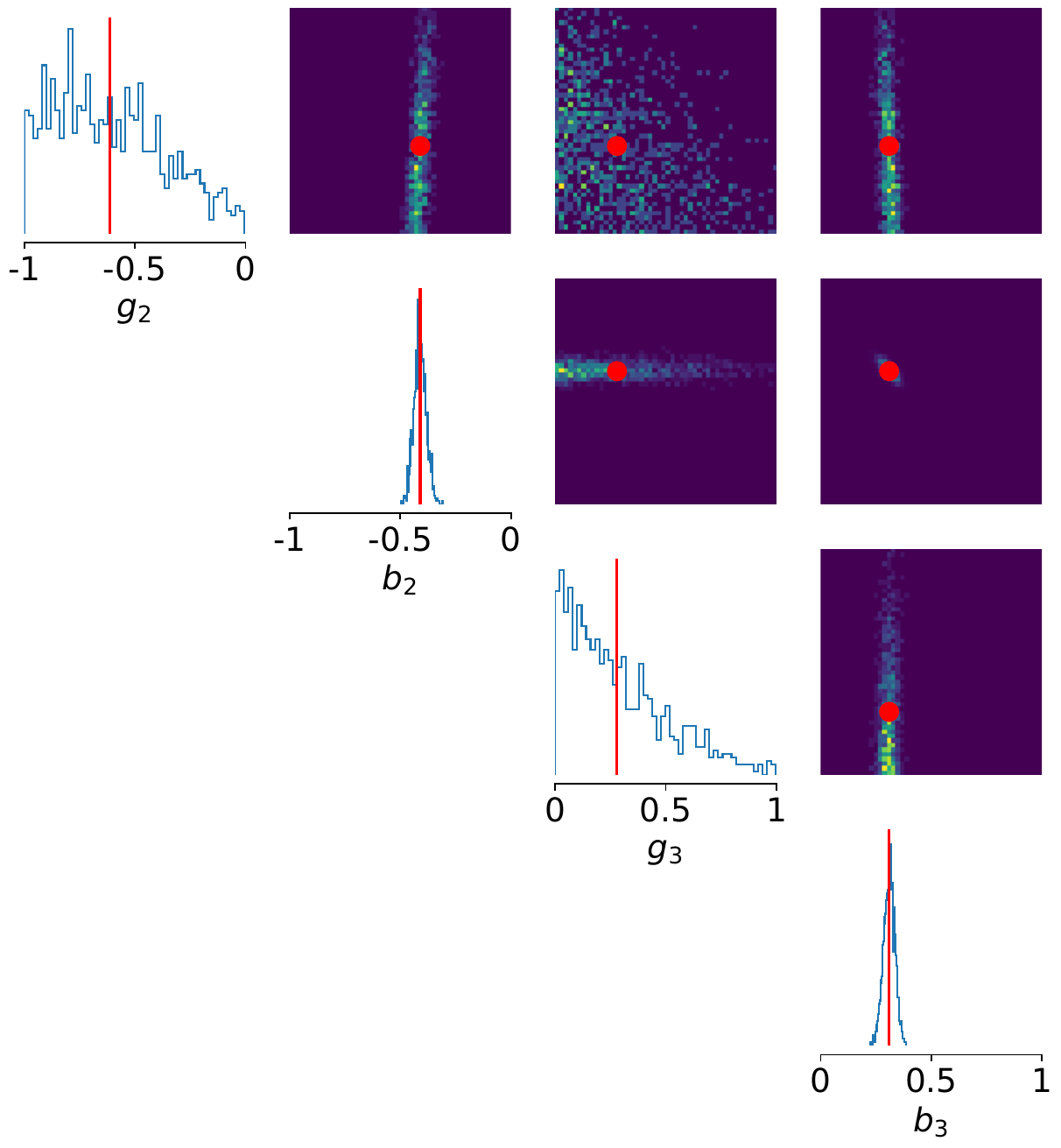}
\caption{Approximate ground truth}
\end{subfigure}
\hfill
\begin{subfigure}[t]{0.32\textwidth}\centering
\includegraphics[width=\linewidth]{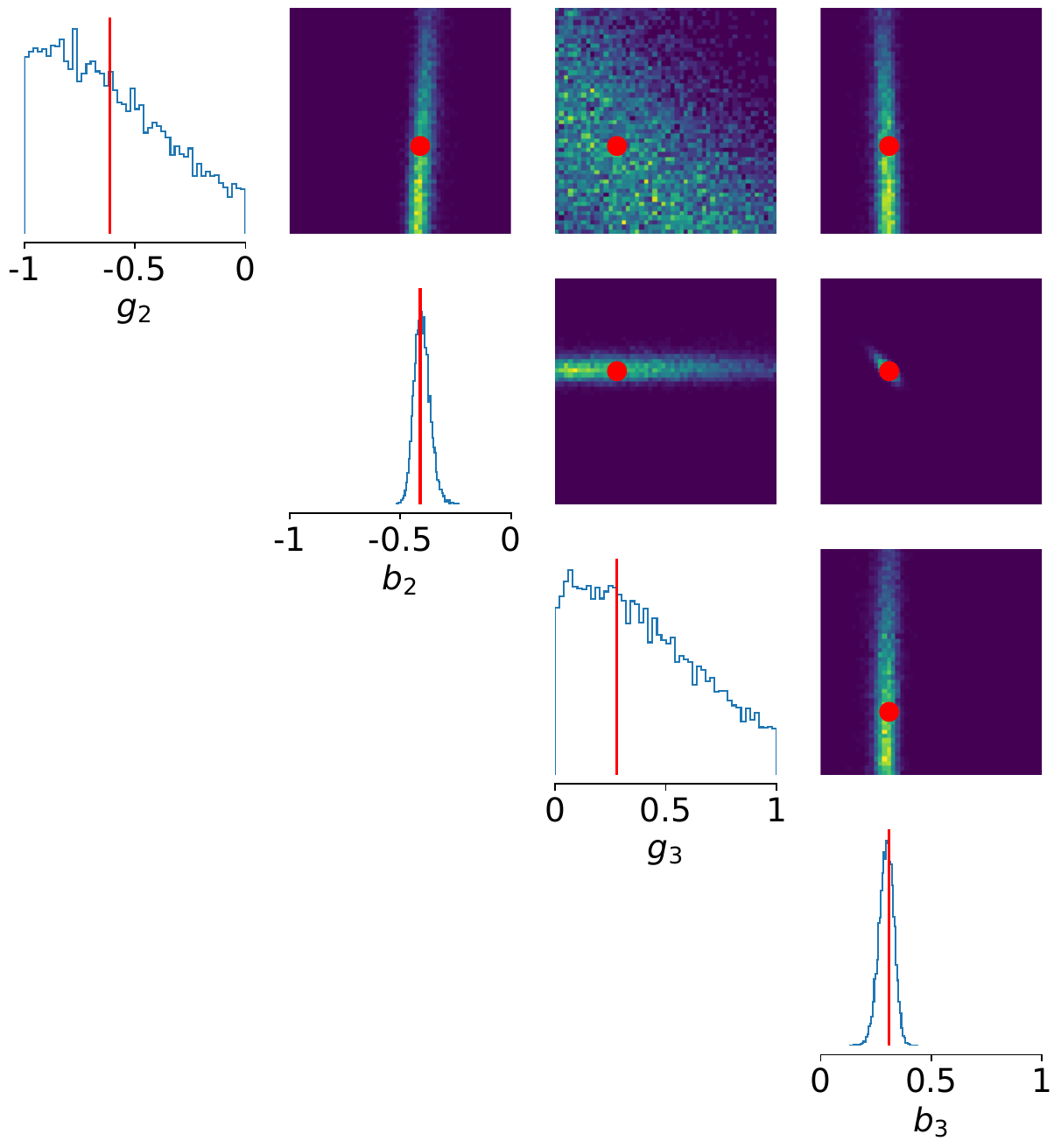}
\caption{SNPE}
\end{subfigure}
\hfill
\begin{subfigure}[t]{0.32\textwidth}\centering
\includegraphics[width=\linewidth]{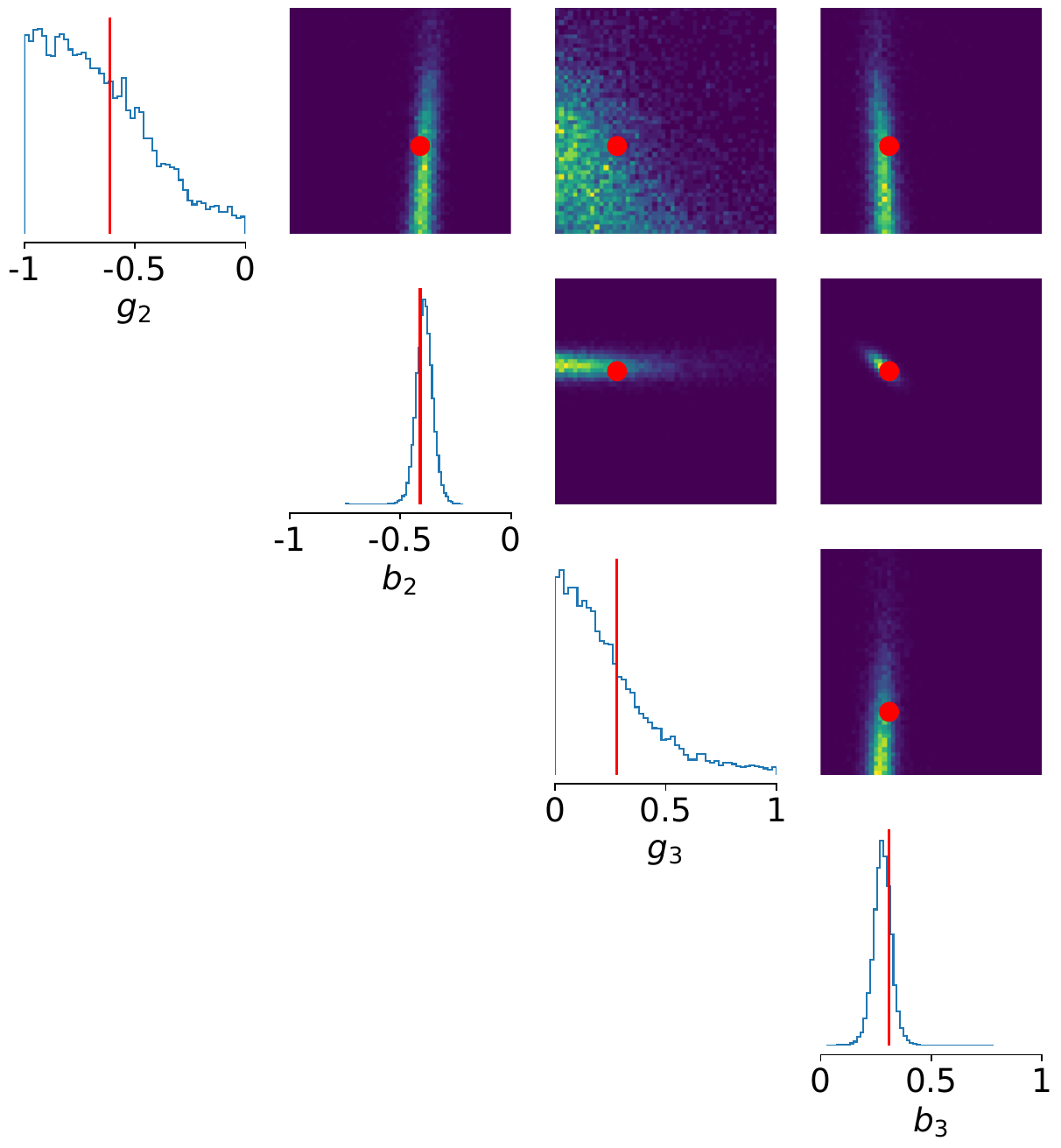}
\caption{DF-NPE}
\end{subfigure}
\caption{Posterior samples from the approximate ground truth, SNPE and DF-NPE for the noisy Brock--Hommes setting. DF-NPE uses Mixed training at a directly-simulated:synthetic ratio of 3:7. Diagonal panels show marginal posterior distributions as count histograms; upper-triangular panels show bivariate joint distributions as two-dimensional count histograms. Red lines and points denote the mean of the likelihood-based reference posterior. Darker and lighter colours indicate lower and higher sample counts, respectively, in the off-diagonal plots.}
\label{fig:bh_noisy_overview}
\end{figure}

\paragraph{Noisy setting.}\label{sec:bh_noisy}
In the noisy setting, the reference posterior has broad marginals for $g_2$ and $g_3$ but concentrated marginals for $b_2$ and $b_3$ (Figure~\ref{fig:bh_noisy_overview}). Both displayed estimators reproduce this contrast, which appears as narrow bands in the joint distributions of trend coefficients and biases. They differ more visibly in $g_3$, where Mixed places more mass towards the lower end of the support than SNPE. The bias marginals remain concentrated in both fits, although their locations and spreads differ from the reference. Recovery in this setting therefore involves reproducing both the broad trend-coefficient distributions and the narrow bias distributions.

\begin{figure}[!htbp]
\centering
\begin{subfigure}[t]{0.96\textwidth}\centering
\includegraphics[width=\linewidth]{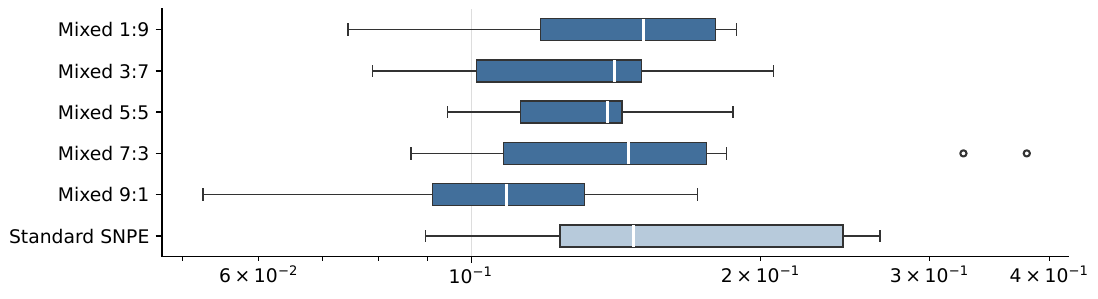}
\caption{Wasserstein distance.}\end{subfigure}\par\vspace{0.4em}
\begin{subfigure}[t]{0.96\textwidth}\centering
\includegraphics[width=\linewidth]{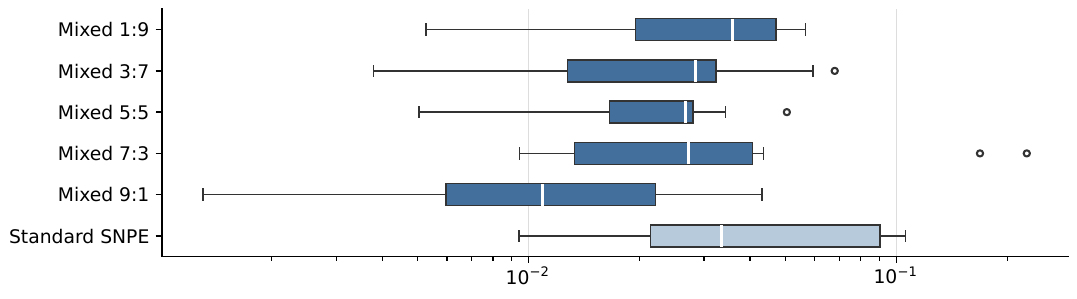}
\caption{Squared maximum mean discrepancy.}\end{subfigure}
\caption{Posterior discrepancies of SNPE and DF-NPE with Mixed training for the noisy Brock--Hommes setting. Results show WASS and MMD$^2$ over ten fits per configuration using one fixed observation. Ratios denote directly-simulated:synthetic allocations; SNPE uses only direct simulations. White lines denote medians, boxes show the interquartile range, and whiskers extend to the most extreme values within 1.5 interquartile ranges.}
\label{fig:noisy_fidelity}
\end{figure}

Across repeated fits, the Mixed and SNPE discrepancy distributions overlapped more than in the smooth setting (Figure~\ref{fig:noisy_fidelity}). Mixed 9:1 had the lowest median for both metrics, while 3:7 and 5:5 also had lower medians than SNPE. The strongest simulation saving occurred at 1:9, but neither median improved on the baseline. Its mean discrepancies were nevertheless lower (Table~\ref{tab:historical_all}), so the 90\% reduction in direct simulations was accompanied by improved average scores rather than lower median scores.

A different mean--median contrast appeared at 7:3. Its central scores were comparable to those of the other intermediate allocations, but both discrepancy distributions contained high-discrepancy outliers. In particular, its median MMD$^2$ was lower than that of SNPE, while its mean was higher. These results explain why increasing the directly-simulated fraction cannot be interpreted as a steady improvement in posterior recovery.

Taken together, the allocation experiments identify 1:9 as the largest saving in direct simulations among the tested configurations, with lower mean discrepancies in both settings. The allocations giving the lowest median discrepancies differed, however: 5:5 in the smooth setting and 9:1 in the noisy setting. The choice of allocation therefore depends on the simulation budget and the posterior accuracy required for the task.

\subsection{Further analysis}
\label{sec:further_analysis}
\subsubsection{Comparison of posterior-training strategies}
\label{sec:order_results}\label{sec:ordered_experiments}

The allocation experiments used Mixed training throughout, leaving open whether the same banks are better used in successive stages. We examine this question by comparing Mixed with Synthetic-to-Simulation and Simulation-to-Synthetic at the shared allocations. Figures~\ref{fig:ordered_bh_smooth_accuracy} and~\ref{fig:ordered_bh_noisy_accuracy} show the resulting discrepancy distributions for the two settings.

In the smooth setting, Mixed had lower median discrepancies than either ordered strategy at 1:9 and 3:7. This separation became less clear at 5:5, where the boxes overlapped substantially and WASS and MMD$^2$ gave different median rankings. Across these allocations, Synthetic-to-Simulation moved towards lower discrepancies as the directly-simulated fraction increased, whereas Simulation-to-Synthetic retained broad score distributions. Although ending with directly-simulated data improved some comparisons, it provided no consistent advantage across allocations.

The noisy setting showed a different pattern. Both ordered strategies had lower median discrepancies than Mixed at 1:9, and Synthetic-to-Simulation at 3:7 achieved the lowest medians for both metrics. Its WASS interquartile range was also compact, and WASS was lower than for the reverse order in all ten paired-seed comparisons (Appendix~\ref{app:ordered_results}). This advantage did not extend uniformly to 5:5: Synthetic-to-Simulation retained a low median but also contained high-discrepancy outliers, giving a less favourable mean ranking.

The posterior figure in Appendix~\ref{app:posterior_geometry} show how the ordered fits differ at 3:7. In the smooth setting, both spread $g_2$ more broadly than the reference, while keeping the other marginals concentrated. In the noisy setting, both reproduced the contrast between broad trend-coefficient marginals and narrow bias marginals, with differences in location and tail behaviour. These individual fits illustrate the posterior shapes underlying the discrepancy comparisons, without replacing the evidence across seeds.

The preferred strategy consequently depended on the setting and allocation. This comparison also includes differences beyond source order: Mixed uses early stopping, while each ordered stage uses 500 updates. Since the resulting posterior determines the next proposal, these training choices propagate into later data banks. The results therefore compare complete sequential procedures rather than isolate the effect of the final training source.

\begingroup
\captionsetup{font=footnotesize}
\captionsetup[subfigure]{font=footnotesize}
\begin{figure}[!htbp]
\centering
\begin{subfigure}[t]{0.82\textwidth}
\centering
\includegraphics[width=\linewidth]{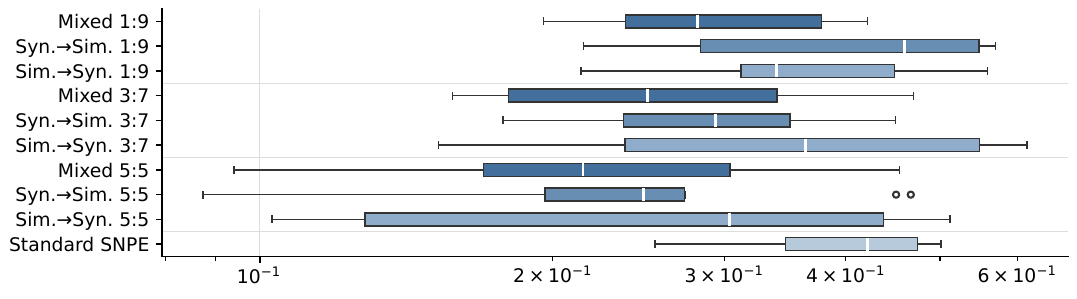}
\caption{Wasserstein distance.}
\end{subfigure}
\par\vspace{0.1em}
\begin{subfigure}[t]{0.82\textwidth}
\centering
\includegraphics[width=\linewidth]{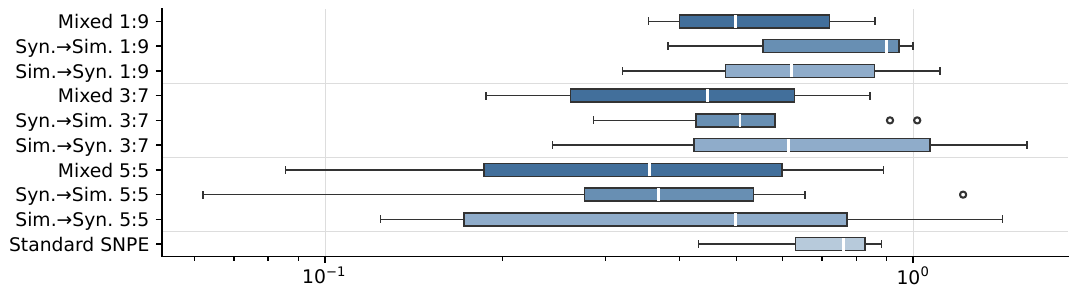}
\caption{Squared maximum mean discrepancy.}
\end{subfigure}
\caption{Posterior discrepancies of the three DF-NPE training strategies and SNPE for the smooth Brock--Hommes setting. Results show WASS and MMD$^2$ over ten fits at directly-simulated:synthetic ratios of 1:9, 3:7 and 5:5. Syn.$\to$Sim. and Sim.$\to$Syn. denote Synthetic-to-Simulation and Simulation-to-Synthetic, respectively. White lines denote medians, boxes show the interquartile range, and whiskers extend to the most extreme values within 1.5 interquartile ranges.}
\label{fig:ordered_bh_smooth_accuracy}
\end{figure}

\begin{figure}[!htbp]
\centering
\begin{subfigure}[t]{0.82\textwidth}
\centering
\includegraphics[width=\linewidth]{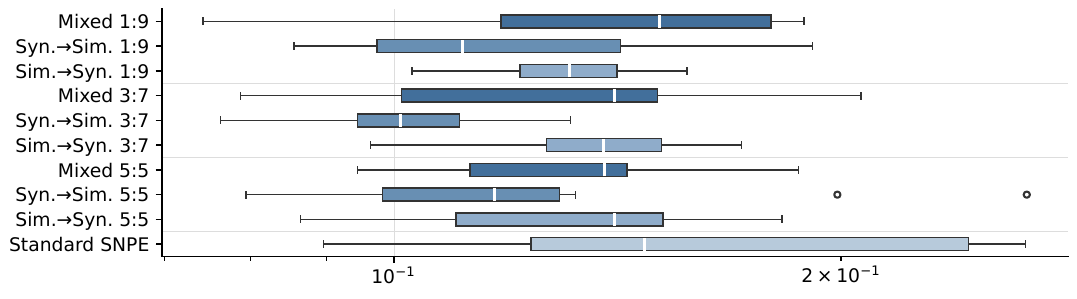}
\caption{Wasserstein distance.}
\end{subfigure}
\par\vspace{0.1em}
\begin{subfigure}[t]{0.82\textwidth}
\centering
\includegraphics[width=\linewidth]{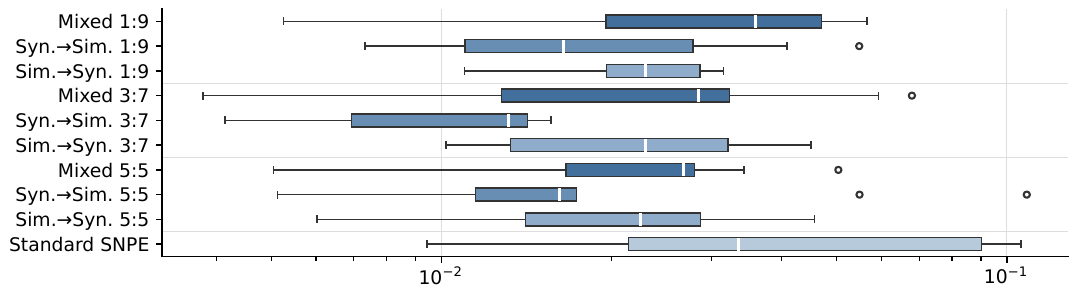}
\caption{Squared maximum mean discrepancy.}
\end{subfigure}
\caption{Posterior discrepancies of the three DF-NPE training strategies and SNPE for the noisy Brock--Hommes setting. Results show WASS and MMD$^2$ over ten fits at directly-simulated:synthetic ratios of 1:9, 3:7 and 5:5. Syn.$\to$Sim. and Sim.$\to$Syn. denote Synthetic-to-Simulation and Simulation-to-Synthetic, respectively. White lines denote medians, boxes show the interquartile range, and whiskers extend to the most extreme values within 1.5 interquartile ranges.}
\label{fig:ordered_bh_noisy_accuracy}
\end{figure}

\endgroup

\subsubsection{Computational cost}
\label{sec:cost}

\begin{figure}[!htbp]
\centering
\begin{subfigure}[t]{0.96\textwidth}
\centering
\includegraphics[width=\linewidth]{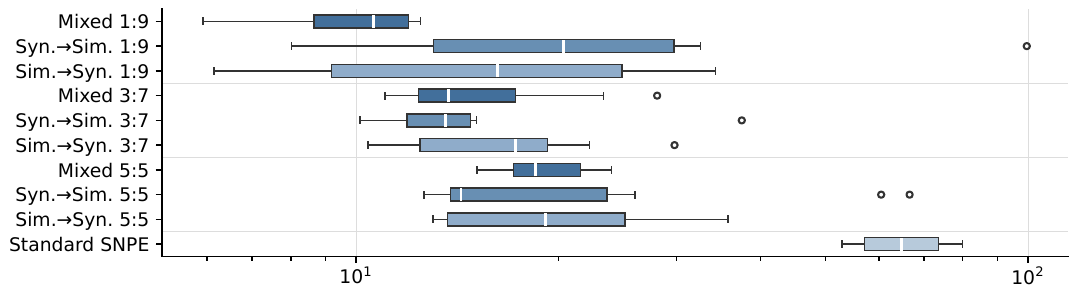}
\caption{Smooth Brock--Hommes.}
\end{subfigure}
\par\vspace{0.4em}
\begin{subfigure}[t]{0.96\textwidth}
\centering
\includegraphics[width=\linewidth]{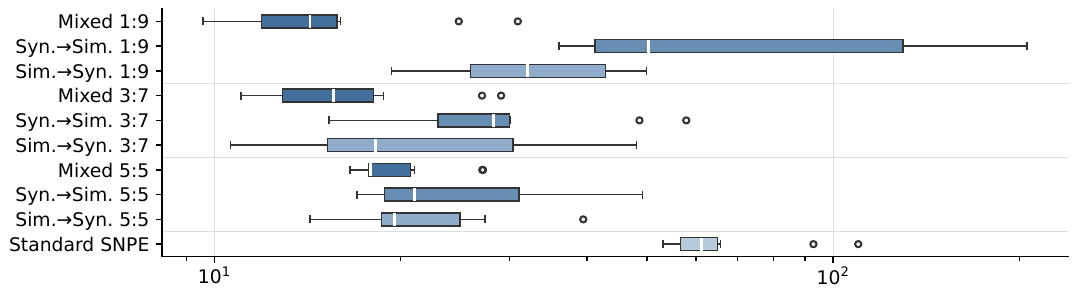}
\caption{Noisy Brock--Hommes.}
\end{subfigure}
\par\vspace{0.4em}

\caption{End-to-end runtime of SNPE and the three DF-NPE training strategies for the smooth and noisy Brock--Hommes settings. Results show ten runs per configuration at directly-simulated:synthetic ratios of 1:9, 3:7 and 5:5, with runtime measured in minutes. White lines denote medians, boxes show the interquartile range, and whiskers extend to the most extreme values within 1.5 interquartile ranges. }
\label{fig:ordered_runtime}
\end{figure}

The discrepancy improvements must also be considered alongside the time required to obtain them. Figure~\ref{fig:ordered_runtime} shows that Mixed 1:9 had the lowest median runtime in both settings, with all displayed Mixed configurations below the SNPE median. The mean runtimes followed the same comparison with the baseline across all five allocations (Table~\ref{tab:historical_all}). At 1:9, Mixed averaged 10.15 minutes for smooth BH and 15.97 minutes for noisy BH, compared with 65.45 and 67.60 minutes for SNPE.

Ordered training generally required more time and showed greater variation, particularly for Synthetic-to-Simulation at noisy 1:9. At this allocation, its broad runtime distribution contrasted with the shorter Mixed runs despite an identical direct simulation budget. Thus, the lower noisy-setting discrepancies obtained by ordered training came with additional computational cost.

The difference at a shared directly-simulated budget reflects the other costs included in each workflow, including encoder fitting, surrogate fitting and posterior training. Synthetic generation also recurs after every encoder update, producing $5.5N_s$ accepted pairs over ten rounds rather than only the final $N_s$ pairs. Together with the neural-training costs shown in Figure~\ref{fig:runtime_phases}, this repeated generation contributes to the measured runtime. A 90\% reduction in direct simulations at 1:9 therefore need not produce a proportional reduction in total time.

\subsubsection{Surrogate and augmentation diagnostics}
\label{sec:offline_diagnostics}

The preceding comparisons establish the performance of DF-NPE as a complete procedure, but several components change together. To examine the contribution of augmentation under a fixed representation, we held the encoder, surrogate, directly-simulated bank and posterior initialisation fixed. Both control arms used 1,000 posterior updates; the augmented arm additionally received synthetic pairs. Because these pairs replaced some directly-simulated data exposure within that budget, the comparison measures the effect of augmentation under a fixed update budget.

Under these controls, augmentation affected posterior recovery differently across settings. With Synthetic-to-Simulation sources at 1:9, it lowered diagnostic MMD$^2$ in nine of ten smooth-setting pairs but worsened eight noisy-setting pairs. The surrogate checks also found positive mean excess discrepancies in every tested source configuration, indicating remaining differences between generated and directly encoded summaries. These findings help explain why the benefits of the complete procedure cannot be attributed to synthetic pairs alone. Appendix~\ref{app:offline_diagnostics} reports the full control and surrogate results; predictive checks are reported in Appendix~\ref{app:predictive_scores}.

\FloatBarrier

\section{Discussion and Conclusion}
\label{sec:discussion}\label{sec:conclusion}

DF-NPE reduces the demand for direct simulations by generating additional training pairs in a learned summary space. Direct simulations update this space through the encoder and train the conditional surrogate, which then supplies synthetic summaries for posterior estimation. In the two Brock--Hommes settings, this procedure recovered posterior structure with fewer direct simulations and lower recorded runtimes than SNPE. These experiments extend the simulation-efficiency comparisons of \citet{DyerEtAl2024} to the composition of the data used for neural posterior training.

The clearest saving occurred with Mixed 1:9, which used 1,000 direct simulations against 10,000 for SNPE. Its mean WASS and MMD$^2$ were lower in both settings, although its noisy-setting medians did not improve on the baseline. Other allocations achieved lower discrepancies, so the largest simulation saving and the best posterior agreement occurred at different allocations. This distinction matters when selecting a budget: the repeated-fit distributions show the accuracy and variability associated with each choice, while the posterior figures reveal the corresponding distributional structure.

The way these pairs were used also affected recovery. Mixed performed favourably in the smooth setting, whereas Synthetic-to-Simulation at noisy 3:7 achieved the lowest median discrepancies among the tested strategies. A strategy that works well for one setting therefore need not be preferred in another. Moreover, each fitted posterior determines the next proposal, allowing training choices to affect subsequent data as well as the current fit. Together with the different stopping rules, this feedback explains why the strategy comparisons concern complete inference procedures.

The frozen-component controls provide a narrower view of synthetic augmentation within those procedures. Once the representation and surrogate were fixed, adding generated pairs improved some configurations and worsened others. This dependence on the fitted components is consistent with the augmented distribution in Appendix~\ref{app:implementation}, which contains both encoded simulator output and surrogate-generated summaries. Proposal correction adjusts for parameter sampling but does not remove differences between these conditional distributions. The observed gains therefore support evaluating augmentation as part of DF-NPE, with its allocation chosen according to posterior recovery.

That choice also determines computational cost. Mixed 1:9 had the lowest recorded runtime in both settings, while ordered training could take substantially longer at the same directly-simulated budget. The additional time includes re-encoding, surrogate fitting, repeated synthetic generation and posterior training. Evaluating direct calls together with total runtime consequently shows whether a reduction in simulation demand translates into a faster calibration procedure.

The present evidence comes from repeated fits to one simulated observation per setting, assessed against finite MCMC reference samples. Extending the evaluation to independent observations would allow simulation-based calibration to assess uncertainty more broadly \citep{TaltsEtAl2018,HermansEtAl2022}. Further ABM families would test the transfer of the allocation and strategy results, while applications to market data would also require checks for simulator misspecification \citep{WardEtAl2022,SchmittEtAl2024}.

Within the tested Brock--Hommes settings, DF-NPE shows that generated latent summaries can supplement a limited directly-simulated training set while retaining useful posterior information. Its practical value lies in making that substitution available within sequential posterior estimation, with the allocation and training strategy assessed through posterior recovery and computational cost.

\FloatBarrier
\bibliographystyle{apalike}
\bibliography{references}

\clearpage
\appendix
\numberwithin{figure}{section}
\numberwithin{table}{section}
\numberwithin{equation}{section}

\clearpage
\section{Implementation details of DF-NPE}\label{app:implementation}
\subsection{Architecture and optimisation}
Table~\ref{tab:common_training_budget} gives the shared architecture and Mixed optimisation settings. Neural network parameters are updated with Adam \citep{KingmaBa2015}.
\begin{table}[H]
\centering
\caption{Architecture and optimisation settings for Mixed and SNPE. Sequential budgets and allocations are specified in Section~\ref{sec:evaluation}.}
\label{tab:common_training_budget}
\footnotesize
\begin{tabular}{p{0.28\textwidth}p{0.64\textwidth}}
\toprule
Component & Setting \\
\midrule
Supervised encoder & Two-layer GRU, hidden width 32 and latent dimension 8; parameter-regression head with one 32-unit hidden layer \\
Encoder fit & Adam, learning rate $10^{-3}$, batch size 100, at most 100 epochs; latent extraction batch size 256 \\
Conditional latent surrogate & Five-transform MAF with 50 hidden features; learning rate $5\times10^{-4}$, batch size 100, 10\% validation and early stopping after 10 validation epochs \\
Latent posterior estimator & Five-transform SNPE-C MAF with 50 hidden features; learning rate $5\times10^{-4}$, batch size 200, 10\% validation and early stopping after 20 validation epochs \\
SNPE comparator & Jointly trained two-layer GRU embedding and MAF posterior through the project defaults \\
Execution & Timing runs capped neural fits at 500 epochs \\
\bottomrule
\end{tabular}

\end{table}

For ordered training, the surrogate settings in Table~\ref{tab:common_training_budget} use a maximum of 100 epochs. Posterior stages use Adam at $5\times10^{-4}$ and the update counts and batch size in Section~\ref{sec:evaluation}. Incomplete minibatches are dropped, and shuffled data loaders are cycled until the required number of updates is reached. Post-sequential fine-tuning is disabled.

\label{app:display_sampling}Timing runs use one simulator worker on macOS 15.6.1 ARM64 with Python 3.12.7; timing boundaries follow Section~\ref{sec:metrics}.

\subsection{Source sampling and comparison controls}\label{app:schedules}
The source-sampling distribution is defined in Equation~\eqref{eq:sampling_schedule}, with the three schedules in Equations~\eqref{eq:mixed_schedule}--\eqref{eq:simulation_to_synthetic}. Mixed uses $\lambda_{k,t}=\rho_k=n_r^{(k)}/(n_r^{(k)}+n_s^{(k)})$ throughout; ordered training switches from 0 to 1 or from 1 to 0 after 500 updates. This specifies source sampling, not a decomposition of the atomic SNPE-C loss: pooled minibatches can contain contrastive atoms from both sources.

\subsection{Meaning of the augmented distribution}
Proposal correction does not correct surrogate error. If $p_{\omega_k}(h\mid\theta)$ is the encoded simulator distribution, a constant directly-simulated fraction $\rho$ gives the population mixture
\begin{equation}
\overline p_{k,\rho}(h\mid\theta)=\rho p_{\omega_k}(h\mid\theta)+(1-\rho)q_{\phi_k}(h\mid\theta).
\label{eq:effective_conditional}
\end{equation}
Under adequate proposal support and ideal density estimation, the target is proportional to $p(\theta)\overline p_{k,\rho}(h\mid\theta)$. This expression characterises the training target, without establishing convergence or eliminating representation, surrogate and density-estimation errors.

\section{Further experimental details}
\label{app:models}

\subsection{Scoring details and predictive checks}\label{app:score_details}

Each configuration uses seeds 123, 456, 789, 101, 202, 303, 404, 505, 606 and 707 for the same observation. Budgets and allocations are specified in Section~\ref{sec:evaluation}.

The six historical smooth-regime seed-123 configurations contain 10,000 draws: distributional scoring uses the archived 1,000-draw subsample, while MAE retains the full mean. Other numerical sample sets contain 1,000 draws. The MAE estimates therefore have unequal Monte Carlo precision in this historical subset.

\subsubsection{Posterior-mean parameter error}

For the known generating parameter $\theta^*$, we report the mean absolute error of the estimated posterior mean,
\begin{equation}
\operatorname{MAE}_{\theta^*}
=\frac{1}{d_\theta}\sum_{\ell=1}^{d_\theta}
|\bar\vartheta_\ell-\theta^*_\ell|,
\qquad \bar\vartheta=\frac{1}{m_\mu}\sum_{j=1}^{m_\mu}\vartheta_j,
\label{eq:evaluation_mae}
\end{equation}
where $d_\theta=4$ and $m_\mu$ is the number of draws used to estimate the mean. Appendix~\ref{app:score_details} documents the historical sample-count exception for this statistic. Lower MAE indicates a posterior mean closer to the generating coefficients. It does not establish accurate posterior uncertainty: the reference posterior mean itself need not equal $\theta^*$ for a finite observed trajectory. Accordingly, MAE is secondary to the two distributional discrepancies when assessing posterior recovery \citep{LueckmannEtAl2021}.

\subsubsection{Posterior predictive scores}
\label{app:predictive_scores}

Posterior predictive checks assess model fit by comparing observations with replicated data drawn under the posterior \citep{GelmanEtAl1996}. For smooth Brock--Hommes, let $x_t^{(r)}$, $r=1,\ldots,R_{\mathrm{pp}}$, denote trajectories simulated from $R_{\mathrm{pp}}=200$ posterior draws per fit. For the observed trajectory $y_{o,t}$, let $q_{a,t}$ be the empirical $a$-quantile of the simulated values at time $t$. The reported pointwise-median RMSE and observed-point coverage are
\begin{equation}
\begin{aligned}
\operatorname{RMSE}_{\mathrm{med}}&=\left[\frac{1}{T_o}\sum_{t=1}^{T_o}(q_{0.5,t}-y_{o,t})^2\right]^{1/2},\\
C_{0.9}&=\frac{1}{T_o}\sum_{t=1}^{T_o}\mathbf 1\{q_{0.05,t}\le y_{o,t}\le q_{0.95,t}\},
\end{aligned}
\end{equation}
where $T_o$ is the stored trajectory length. The six-summary error uses $s(x)$ containing the trajectory mean, standard deviation, minimum, maximum, last-minus-first value and correlation with a linear time index. The standard deviation uses divisor $T_o$; the time correlation is set to zero for a constant trajectory. For each component, we compute the median across predictive trajectories and report
\begin{equation}
\operatorname{MAE}_{\mathrm{summary}}
=\frac{1}{6}\sum_{\ell=1}^{6}
\left|\operatorname{median}_{r}s_\ell(x^{(r)})-s_\ell(y_o)\right|.
\end{equation}
These summaries are not rescaled before averaging, so their contributions depend on their numerical scales. This trajectory-summary error is distinct from $\operatorname{MAE}_{\theta^*}$, which concerns parameter means. Lower RMSE and summary MAE indicate closer predictive agreement under these particular losses. Coverage is the fraction of observed time points inside pointwise bands, not repeated-observation parameter coverage or simultaneous trajectory coverage. Because interval widths are unavailable, larger coverage alone cannot establish better predictive performance.

Table~\ref{tab:ppc} gives the complete predictive scores. Some errors are larger for the finite reference sample than for fitted estimators; this unresolved discrepancy does not establish superior posterior recovery by a fitted method.

\begin{table}[htbp]
\centering\small
\caption{Smooth Brock--Hommes: posterior-predictive checks over ten fits, using 200 posterior draws per fit returned to the simulator. Entries report mean $\pm$ sample standard deviation.}
\label{tab:ppc}
\begin{tabular}{lccc}
\toprule
Workflow & 90\% coverage & Median RMSE & Summary MAE \\
\midrule
1:9 & $0.979 \pm 0.020$ & $0.577 \pm 0.294$ & $0.380 \pm 0.318$ \\
3:7 & $0.931 \pm 0.093$ & $0.459 \pm 0.074$ & $0.267 \pm 0.105$ \\
5:5 & $0.957 \pm 0.050$ & $0.513 \pm 0.060$ & $0.190 \pm 0.027$ \\
7:3 & $0.951 \pm 0.097$ & $0.482 \pm 0.089$ & $0.227 \pm 0.063$ \\
9:1 & $0.973 \pm 0.022$ & $0.484 \pm 0.083$ & $0.203 \pm 0.049$ \\
SNPE & $0.932 \pm 0.011$ & $0.372 \pm 0.015$ & $0.222 \pm 0.014$ \\
Reference sample & $0.924 \pm 0.012$ & $0.656 \pm 0.041$ & $0.522 \pm 0.028$ \\
\bottomrule
\end{tabular}
\end{table}

\FloatBarrier

\subsubsection{Posterior display conventions}

Posterior displays use the \texttt{pairplot} implementation of \citet{DyerEtAl2024}, with 50-bin marginals, $50\times50$ bivariate bins over task-specific prior ranges and blank lower triangles. Sample counts, scaling and seed selection follow Section~\ref{sec:evaluation}.

\subsection{Score harmonisation and provenance}

The original ordered-suite diagnostics use mean marginal Wasserstein distances, a biased MMD$^2$ estimator with a squared-distance kernel scale, and posterior-mean error relative to the reference mean. For comparison with Mixed, all 120 ordered final sample sets are rescored with the joint Euclidean Wasserstein cost, unbiased MMD$^2$ and error relative to the generating parameter defined in Section~\ref{sec:metrics}. With equally weighted sample sets of the same size, joint Wasserstein is computed as the minimum-cost assignment divided by the sample count, which is equivalent to the empirical optimal-transport definition here.

For the six smooth Brock--Hommes seed-123 exceptions described above, seed $20260828+123$ selects the 1,000-draw WASS and MMD$^2$ subsample. Ordered provenance retains source paths, run fingerprints and sample hashes.

\section{Supplementary posterior and computational results}
\label{app:additional_results}
\label{app:posterior_geometry}\label{app:ordered_posteriors}\label{app:bh_noisy}

This appendix brings together posterior geometry, complete numerical comparisons and runtime breakdowns for the workflows evaluated in Section~\ref{sec:experiments}.

\subsection{Posterior geometry}
The ordered-strategy figure complements the approximate ground truth, SNPE and DF-NPE (Mixed, 3:7) displays in Figures~\ref{fig:bh_smooth_overview} and~\ref{fig:bh_noisy_overview}.

\begin{figure}[!htbp]
\centering
\begin{subfigure}[t]{0.48\textwidth}\centering
\includegraphics[width=\linewidth,height=0.27\textheight,keepaspectratio]{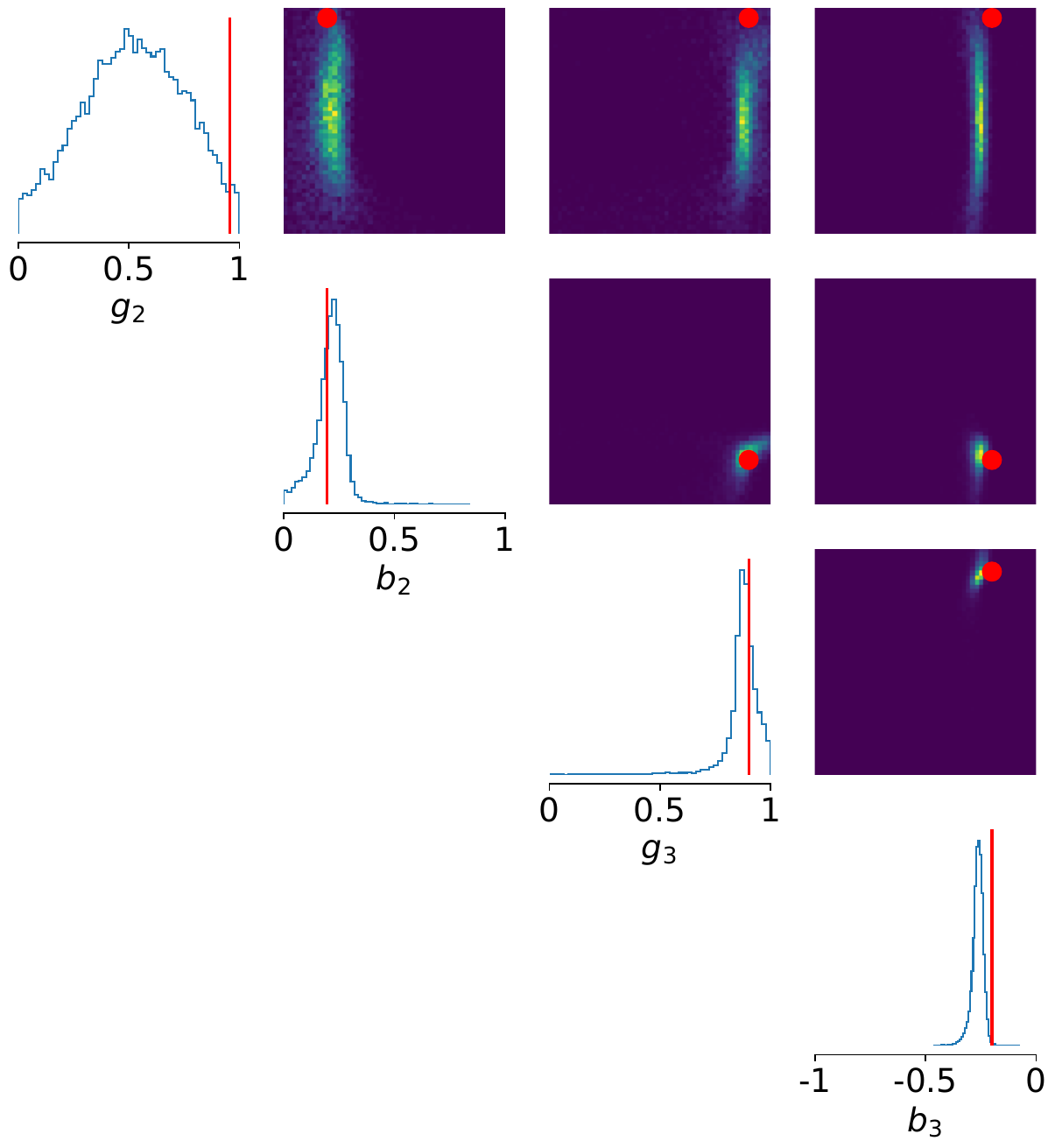}
\caption{Smooth BH: Syn.$\to$Sim.}
\end{subfigure}
\hfill
\begin{subfigure}[t]{0.48\textwidth}\centering
\includegraphics[width=\linewidth,height=0.27\textheight,keepaspectratio]{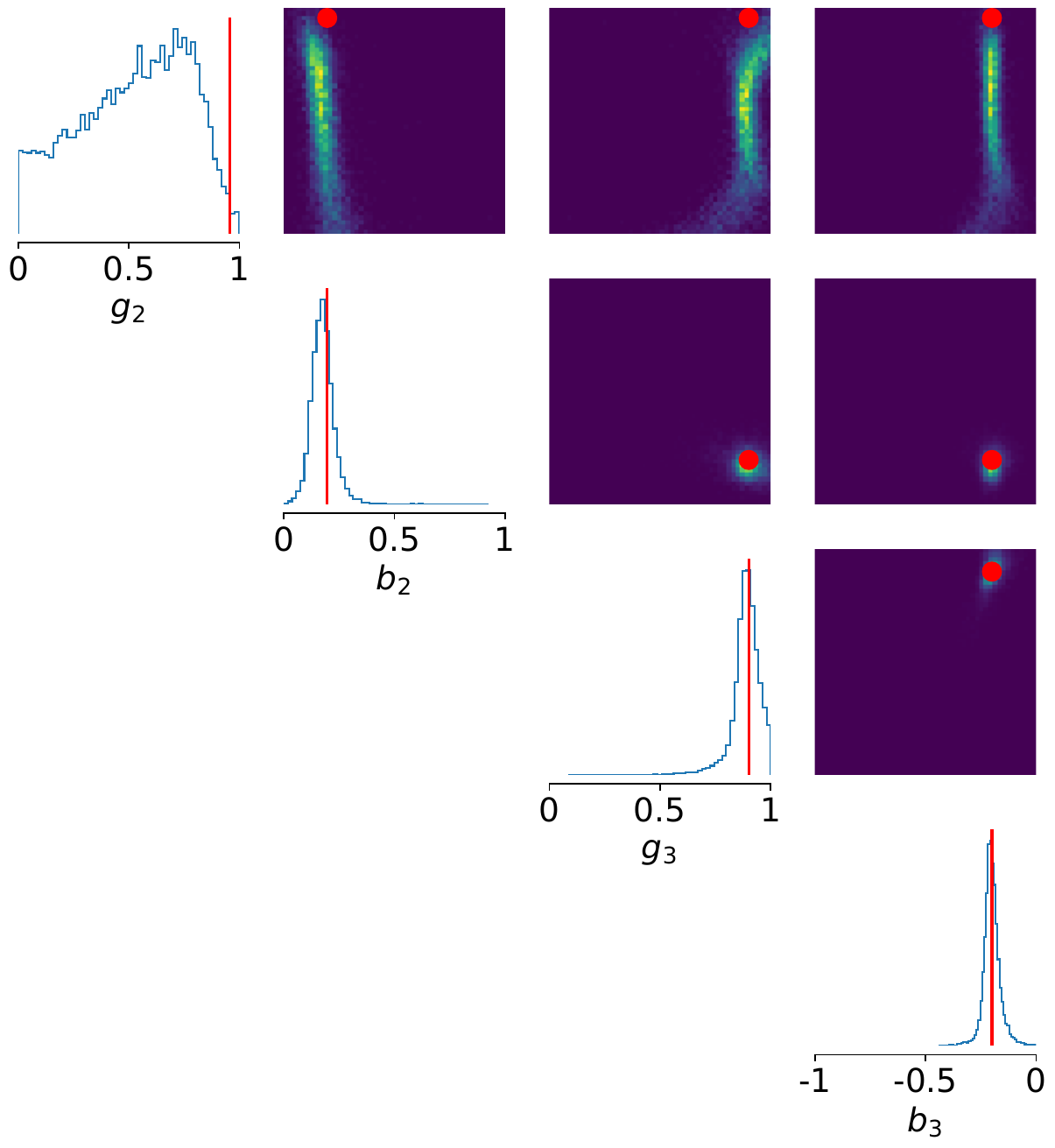}
\caption{Smooth BH: Sim.$\to$Syn.}
\end{subfigure}
\par\medskip
\begin{subfigure}[t]{0.48\textwidth}\centering
\includegraphics[width=\linewidth,height=0.27\textheight,keepaspectratio]{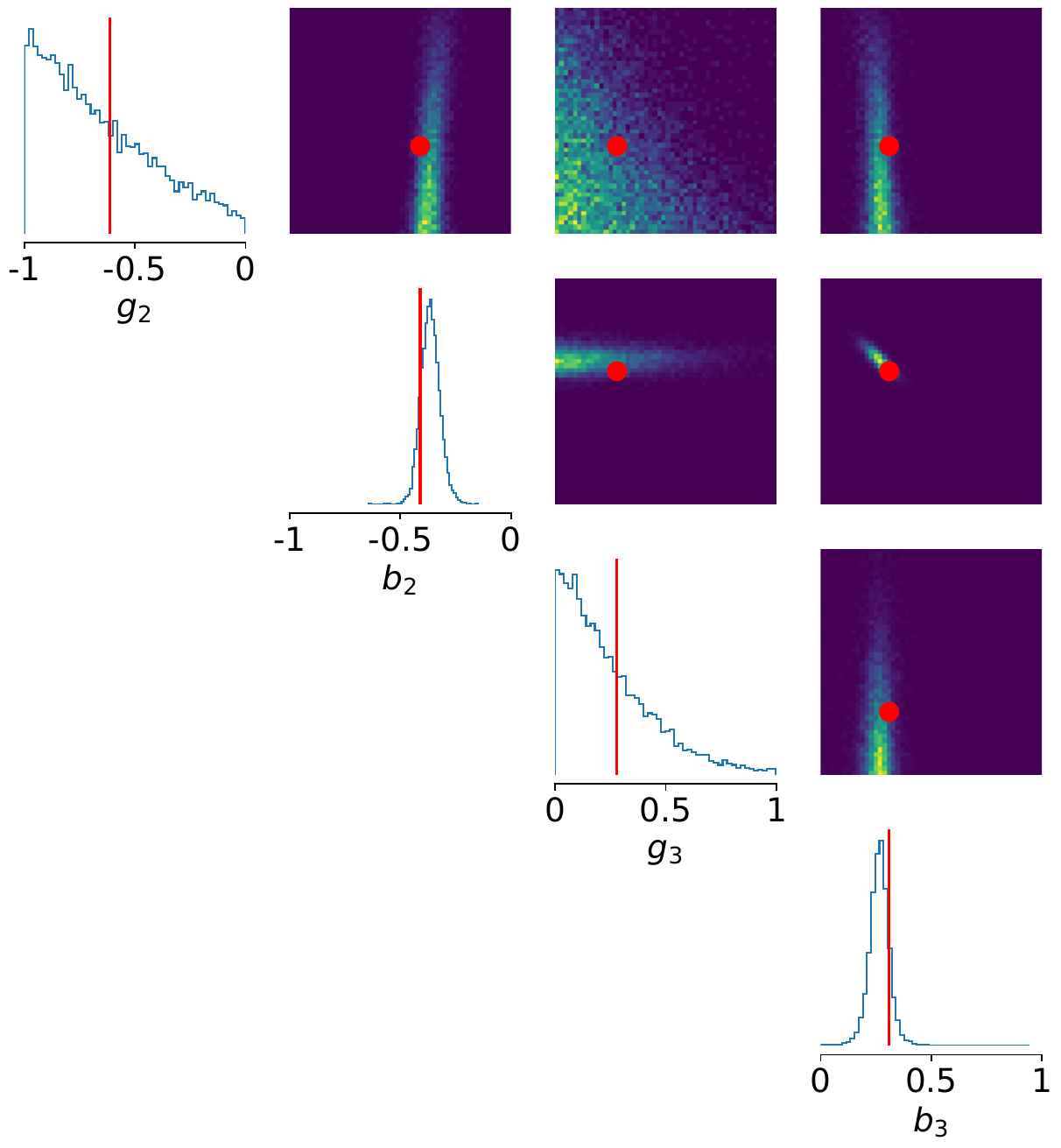}
\caption{Noisy BH: Syn.$\to$Sim.}
\end{subfigure}
\hfill
\begin{subfigure}[t]{0.48\textwidth}\centering
\includegraphics[width=\linewidth,height=0.27\textheight,keepaspectratio]{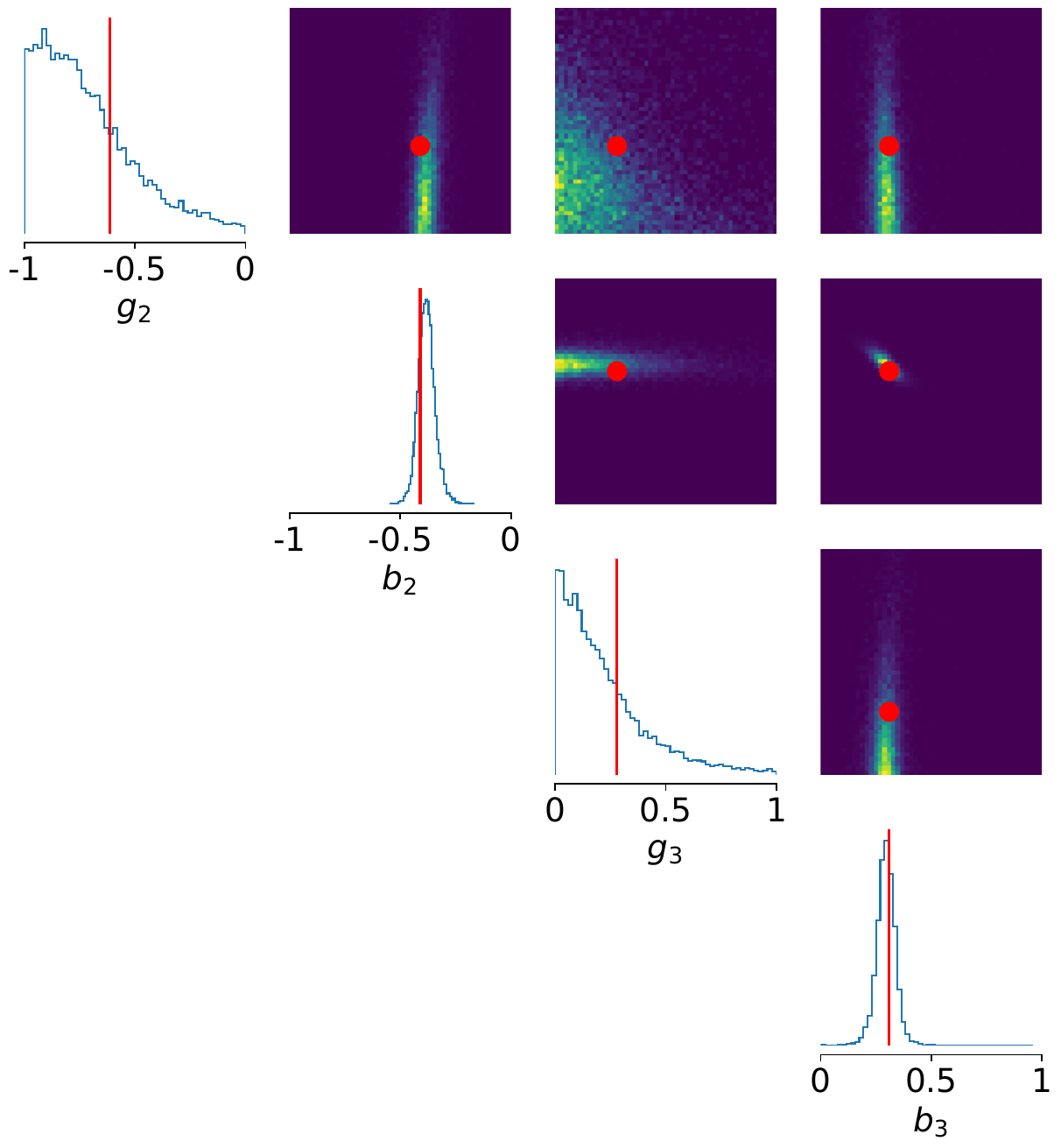}
\caption{Noisy BH: Sim.$\to$Syn.}
\end{subfigure}
\par\medskip
\caption{Posterior samples from the two ordered DF-NPE training strategies for the smooth and noisy Brock--Hommes settings at a directly-simulated:synthetic ratio of 3:7. Syn.$\to$Sim. and Sim.$\to$Syn. denote Synthetic-to-Simulation and Simulation-to-Synthetic, respectively. Diagonal panels show marginal posterior distributions as count histograms; upper-triangular panels show bivariate joint distributions as two-dimensional count histograms. Red lines and points denote the mean of the likelihood-based reference posterior. Darker and lighter colours indicate lower and higher sample counts, respectively, in the off-diagonal plots.}
\label{fig:ordered_bh_smooth_joint}\label{fig:ordered_bh_noisy_joint}
\end{figure}

\subsection{Comparisons at equal direct simulation budgets}
\label{app:matched_budget}
Table~\ref{tab:matched} compares Mixed with SNPE using the same direct simulation budget in the smooth setting. Five paired seeds are available for each allocation. The two procedures retain their respective representation-learning and training protocols.

\begin{table}[htbp]
\centering\small
\caption{Smooth Brock--Hommes: same-direct-budget comparisons over five paired seeds. Entries report mean $\pm$ sample standard deviation of SNPE-minus-Mixed discrepancies; positive values favour Mixed. Parentheses give improving seeds out of five.}
\label{tab:budget_predictive}\label{tab:matched}
\begin{tabular}{lrrrr}
\toprule
Ratio & Direct/round & WASS gain & MMD$^2$ gain & MAE gain \\
\midrule
1:9 & 100 & $0.202 \pm 0.140$ (5/5) & $0.295 \pm 0.308$ (3/5) & $0.083 \pm 0.049$ (5/5) \\
3:7 & 300 & $0.216 \pm 0.096$ (5/5) & $0.403 \pm 0.195$ (5/5) & $0.067 \pm 0.027$ (5/5) \\
5:5 & 500 & $0.229 \pm 0.172$ (5/5) & $0.465 \pm 0.347$ (5/5) & $0.070 \pm 0.054$ (5/5) \\
7:3 & 700 & $0.260 \pm 0.063$ (5/5) & $0.484 \pm 0.150$ (5/5) & $0.076 \pm 0.027$ (5/5) \\
9:1 & 900 & $0.310 \pm 0.057$ (5/5) & $0.573 \pm 0.169$ (5/5) & $0.087 \pm 0.015$ (5/5) \\
\bottomrule
\end{tabular}
\end{table}

\subsection{Mixed allocations and SNPE}
\begin{table}[H]
\centering\small
\setlength{\tabcolsep}{3pt}
\caption{Mixed workflows and SNPE across the two Brock--Hommes settings. Entries are mean $\pm$ sample standard deviation over ten seeds. Scores and timing conventions follow Section~\ref{sec:metrics}. Ratios are directly-simulated:synthetic allocations; time is in minutes.}
\label{tab:historical_all}\label{tab:main_metrics}\label{tab:runtime}\label{tab:bh_noisy_metrics}
\begin{tabular}{lccccc}
\toprule
Workflow & WASS & MMD$^2$ & MAE to $\theta^*$ & Time (min) & Speed-up \\
\midrule
\multicolumn{6}{l}{\textbf{Smooth Brock--Hommes}}\\
1:9 & $0.3051 \pm 0.0841$ & $0.5637 \pm 0.1997$ & $0.0780 \pm 0.0256$ & $10.15 \pm 2.21$ & $6.45\times$ \\
3:7 & $0.2681 \pm 0.1042$ & $0.4595 \pm 0.2168$ & $0.0685 \pm 0.0326$ & $16.03 \pm 5.59$ & $4.08\times$ \\
5:5 & $0.2458 \pm 0.1228$ & $0.4007 \pm 0.2635$ & $0.0590 \pm 0.0382$ & $19.13 \pm 2.91$ & $3.42\times$ \\
7:3 & $0.2489 \pm 0.1239$ & $0.4478 \pm 0.2968$ & $0.0656 \pm 0.0414$ & $21.47 \pm 2.14$ & $3.05\times$ \\
9:1 & $0.2403 \pm 0.1027$ & $0.4430 \pm 0.2559$ & $0.0622 \pm 0.0332$ & $25.63 \pm 1.11$ & $2.55\times$ \\
SNPE & $0.4046 \pm 0.0808$ & $0.7265 \pm 0.1487$ & $0.1053 \pm 0.0239$ & $65.45 \pm 9.41$ & $1.00\times$ \\
\addlinespace
\multicolumn{6}{l}{\textbf{Noisy Brock--Hommes}}\\
1:9 & $0.1460 \pm 0.0383$ & $0.033966 \pm 0.017558$ & $0.0715 \pm 0.0127$ & $15.97 \pm 6.75$ & $4.23\times$ \\
3:7 & $0.1349 \pm 0.0423$ & $0.028714 \pm 0.021016$ & $0.0806 \pm 0.0177$ & $17.15 \pm 6.22$ & $3.94\times$ \\
5:5 & $0.1340 \pm 0.0274$ & $0.024385 \pm 0.012898$ & $0.0729 \pm 0.0126$ & $20.00 \pm 3.98$ & $3.38\times$ \\
7:3 & $0.1726 \pm 0.1001$ & $0.057256 \pm 0.075828$ & $0.0917 \pm 0.0565$ & $25.86 \pm 9.48$ & $2.61\times$ \\
9:1 & $0.1136 \pm 0.0366$ & $0.015977 \pm 0.013607$ & $0.0740 \pm 0.0233$ & $26.76 \pm 3.34$ & $2.53\times$ \\
SNPE & $0.1761 \pm 0.0694$ & $0.052598 \pm 0.039299$ & $0.0597 \pm 0.0097$ & $67.60 \pm 18.63$ & $1.00\times$ \\
\bottomrule
\end{tabular}

\end{table}
\FloatBarrier
\subsection{Runtime phase breakdown}

\begin{figure}[!htbp]
\centering
\includegraphics[width=0.96\textwidth]{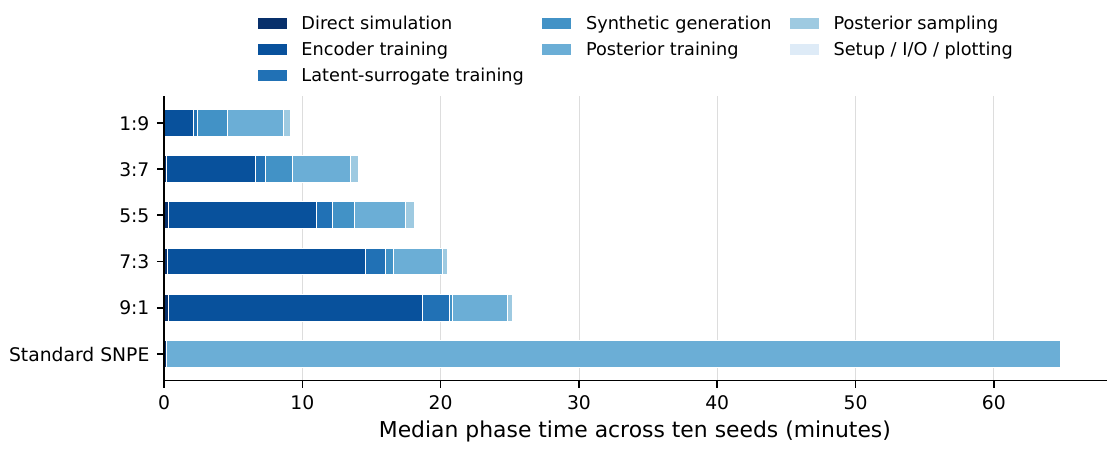}
\caption{Computational cost of SNPE and DF-NPE with Mixed training for the Brock--Hommes model, broken down by workflow phase. Stacked segments show median phase times in minutes over ten seeds; ratios denote directly-simulated:synthetic allocations. Colours identify the computational phases. The setup / I/O / plotting segment contains costs not assigned to the other phases. Segment medians need not sum to the median end-to-end runtime.}
\label{fig:runtime_phases}
\end{figure}
\FloatBarrier
\subsection{Complete ordered-workflow results}
\label{app:ordered_results}

\begin{table}[H]
\centering\footnotesize\setlength{\tabcolsep}{3pt}
\caption{Complete ordered-workflow results across ten seeds (mean $\pm$ sample standard deviation). Syn.$\to$Sim. denotes Synthetic-to-Simulation and Sim.$\to$Syn. denotes Simulation-to-Synthetic. Scores follow Section~\ref{sec:metrics}; time is recorded end-to-end minutes. Mixed and SNPE results are in Table~\ref{tab:historical_all}.}
\label{tab:order_bh_smooth}\label{tab:order_bh_noisy}
\begin{tabular}{llcccc}
\toprule
Ratio & Schedule & WASS & MMD$^2$ & MAE to $\theta^*$ & Time (min) \\
\midrule
\multicolumn{6}{l}{\textbf{Smooth Brock--Hommes}}\\

1:9 & Syn.$\to$Sim. & $0.4182 \pm 0.1407$ & $0.76124 \pm 0.24726$ & $0.1223 \pm 0.0406$ & $27.71 \pm 26.66$ \\
1:9 & Sim.$\to$Syn. & $0.3696 \pm 0.1209$ & $0.67745 \pm 0.27146$ & $0.1017 \pm 0.0376$ & $17.70 \pm 9.69$ \\
\addlinespace
3:7 & Syn.$\to$Sim. & $0.2993 \pm 0.0888$ & $0.55649 \pm 0.23965$ & $0.0804 \pm 0.0258$ & $15.39 \pm 7.99$ \\
3:7 & Sim.$\to$Syn. & $0.3822 \pm 0.1739$ & $0.76860 \pm 0.47356$ & $0.1011 \pm 0.0463$ & $17.22 \pm 5.87$ \\
\addlinespace
5:5 & Syn.$\to$Sim. & $0.2612 \pm 0.1182$ & $0.44882 \pm 0.31734$ & $0.0676 \pm 0.0336$ & $25.23 \pm 20.61$ \\
5:5 & Sim.$\to$Syn. & $0.2932 \pm 0.1624$ & $0.54872 \pm 0.41960$ & $0.0812 \pm 0.0516$ & $20.31 \pm 7.55$ \\
\addlinespace
\multicolumn{6}{l}{\textbf{Noisy Brock--Hommes}}\\

1:9 & Syn.$\to$Sim. & $0.1232 \pm 0.0342$ & $0.02215 \pm 0.01545$ & $0.0656 \pm 0.0165$ & $84.23 \pm 62.70$ \\
1:9 & Sim.$\to$Syn. & $0.1323 \pm 0.0170$ & $0.02316 \pm 0.00657$ & $0.0586 \pm 0.0134$ & $33.52 \pm 10.96$ \\
\addlinespace
3:7 & Syn.$\to$Sim. & $0.1020 \pm 0.0178$ & $0.01095 \pm 0.00460$ & $0.0824 \pm 0.0156$ & $30.20 \pm 13.38$ \\
3:7 & Sim.$\to$Syn. & $0.1388 \pm 0.0223$ & $0.02357 \pm 0.01175$ & $0.0815 \pm 0.0127$ & $24.06 \pm 12.54$ \\
\addlinespace
5:5 & Syn.$\to$Sim. & $0.1326 \pm 0.0573$ & $0.02663 \pm 0.03196$ & $0.0910 \pm 0.0339$ & $26.06 \pm 10.55$ \\
5:5 & Sim.$\to$Syn. & $0.1341 \pm 0.0310$ & $0.02353 \pm 0.01277$ & $0.0849 \pm 0.0346$ & $22.35 \pm 7.11$ \\
\bottomrule
\end{tabular}
\end{table}

\FloatBarrier

Figures~\ref{fig:ordered_bh_smooth_accuracy} and~\ref{fig:ordered_bh_noisy_accuracy} in the main text show the corresponding repeated-fit distributions. The paired comparisons favoured Synthetic-to-Simulation most consistently at noisy 3:7, where WASS improved in all ten pairs and MMD$^2$ in eight. The mean differences (Synthetic-to-Simulation minus Simulation-to-Synthetic) were $-0.0369\pm0.0279$ for WASS and $-0.0126\pm0.0126$ for MMD$^2$ (mean $\pm$ sample standard deviation). At noisy 1:9 and 5:5, WASS improved in six and seven pairs, respectively. In the smooth setting, WASS improved in only four or five pairs per allocation, with both mean discrepancies favouring the reverse order at 1:9. At noisy 5:5, the mean differences favoured Synthetic-to-Simulation for WASS but the reverse order for MMD$^2$. These descriptive comparisons do not identify a uniformly preferred order.

\section{Further diagnostic details and results}
\renewcommand{\arraystretch}{0.96}
\label{app:offline_diagnostics}

The diagnostics use the 120 final encoder--surrogate pairs from the ordered runs in Section~\ref{sec:evaluation}. Historical Mixed sources are excluded because their protocols and replication differ.

Source seeds are the replication unit; sites and trajectories are nested measurements. The diagnostics use final fitted components, shared data banks and one observation per setting.

\subsection{Conditional surrogate fidelity}

For each task, ten parameter sites were drawn from the prior and ten from the finite reference posterior. Reference-posterior sites are used only for offline evaluation; they are unavailable during inference when the reference posterior is unknown. At each site, three independent batches of 100 trajectories supplied calibration, evaluation and direct--direct baseline summaries. The frozen encoder mapped all trajectories to its eight-dimensional representation; the frozen conditional surrogate generated 100 summaries at the same parameter.

Let $\mu_j$ and $s_j$ be coordinate means and sample standard deviations in the direct calibration batch, with floor $\max\{10^{-5}\operatorname{median}_j(s_j),10^{-8}\}$. We standardise all summaries at that site as $z_j=(h_j-\mu_j)/s_j$. The diagnostic kernel is
\begin{equation}
 k_{\mathrm{diag}}(z,z')=\exp\!\left[-\frac{\|z-z'\|^2}{2b}\right],\qquad
 b=\max\{\operatorname{median}_{i<j:\,\|z_i-z_j\|^2>0}\|z_i-z_j\|^2,10^{-8}\},
\end{equation}
where the bandwidth uses the calibration batch; the degenerate all-zero case uses $b=1$. The same unbiased MMD$^2$ estimator is used for direct--synthetic and direct--direct samples. Excess discrepancy is their difference,
\begin{equation}
 E_{\mathrm{sur}}=\widehat{\operatorname{MMD}}^2(z_{\mathrm{test}},z_{\mathrm{syn}})
 -\widehat{\operatorname{MMD}}^2(z_{\mathrm{test}},z_{\mathrm{baseline}}).
\end{equation}
Values are averaged over ten sites within each region and then summarised across ten source fits. Subtracting a finite-sample baseline is descriptive, not a calibrated hypothesis test, and can produce negative values.

\begin{table}[!htbp]
\centering\small\setlength{\tabcolsep}{3pt}
\caption{Conditional surrogate diagnostics over ten source fits (mean $\pm$ source-seed standard deviation). Prior and reference excess MMD$^2$ average ten parameter sites per region, with source-specific standardisation.}
\label{tab:diagnostic_components}\label{tab:diagnostic_surrogate}
\begin{tabular}{lllcc}
\toprule
Task & Source & Ratio & Prior excess MMD$^2$ & Ref. excess MMD$^2$  \\
\midrule
Smooth BH & $G\to S$ & 1:9 & $0.36259\pm0.03831$ & $0.29850\pm0.07948$  \\
Smooth BH & $G\to S$ & 3:7 & $0.22473\pm0.04098$ & $0.15527\pm0.06224$  \\
Smooth BH & $G\to S$ & 5:5 & $0.20284\pm0.03971$ & $0.09657\pm0.02512$  \\
Smooth BH & $S\to G$ & 1:9 & $0.27270\pm0.05115$ & $0.20251\pm0.09400$  \\
Smooth BH & $S\to G$ & 3:7 & $0.25792\pm0.04028$ & $0.13531\pm0.04036$  \\
Smooth BH & $S\to G$ & 5:5 & $0.21534\pm0.04260$ & $0.08688\pm0.03532$  \\
\addlinespace
Noisy BH & $G\to S$ & 1:9 & $0.38752\pm0.06211$ & $0.00999\pm0.00512$  \\
Noisy BH & $G\to S$ & 3:7 & $0.24126\pm0.04675$ & $0.02813\pm0.01664$  \\
Noisy BH & $G\to S$ & 5:5 & $0.22142\pm0.02758$ & $0.02940\pm0.01559$  \\
Noisy BH & $S\to G$ & 1:9 & $0.28685\pm0.02244$ & $0.01083\pm0.00445$  \\
Noisy BH & $S\to G$ & 3:7 & $0.25707\pm0.04170$ & $0.02331\pm0.01211$  \\
Noisy BH & $S\to G$ & 5:5 & $0.22478\pm0.03046$ & $0.03289\pm0.00918$  \\
\bottomrule
\end{tabular}
\end{table}

Site distributions and local standardisation differ between prior and reference regions, so the contrast summarised in Section~\ref{sec:offline_diagnostics} has no causal interpretation. Raw and scale-adjusted density, moment and correlation diagnostics are retained in the source records; density scores do not measure information preservation across representations.

\subsection{Fresh prior-only augmentation controls}

\begin{table}[H]
\centering\scriptsize
\setlength{\tabcolsep}{3pt}
\caption{Frozen-component prior-only controls: augmented-minus-direct differences over ten source fits (mean $\pm$ sample standard deviation). Parentheses give improvements out of ten; negative values favour augmentation. $\Delta\mathrm{MMD}_{\mathrm{diag}}^2$ uses the standardised diagnostic kernel, $\Delta W_{\mathrm{marg}}$ is prior-width-scaled marginal Wasserstein distance, and $\Delta e_{\mathrm{ref}}$ is reference-mean error. Source labels identify the encoder--surrogate origin ($G\to S$: Synthetic-to-Simulation; $S\to G$: Simulation-to-Synthetic). Both augmented arms use pooled training. Both arms share the directly-simulated bank, initialisation and 1,000-update budget, but direct exposure differs. Diagnostic scores use different definitions from the main scores.}
\label{tab:augmentation_summary}\label{tab:diagnostic_posterior}
\begin{tabular}{lllccc}
\toprule
Task & Source & Ratio & $\Delta\mathrm{MMD}_{\mathrm{diag}}^2$ & $\Delta W_{\mathrm{marg}}$ & $\Delta e_{\mathrm{ref}}$ \\
\midrule
Smooth BH & $G\to S$ & 1:9 & $-0.01428\pm0.01305$ (9) & $-0.00933\pm0.01407$ (7) & $-0.00989\pm0.01472$ (7) \\
Smooth BH & $G\to S$ & 3:7 & $-0.00624\pm0.02317$ (5) & $-0.00832\pm0.00733$ (9) & $-0.01047\pm0.00958$ (8) \\
Smooth BH & $G\to S$ & 5:5 & $-0.01409\pm0.01681$ (7) & $-0.01087\pm0.00864$ (9) & $-0.01311\pm0.00914$ (10) \\
Smooth BH & $S\to G$ & 1:9 & $-0.00143\pm0.02758$ (7) & $-0.00077\pm0.01900$ (5) & $-0.00365\pm0.01970$ (5) \\
Smooth BH & $S\to G$ & 3:7 & $-0.00139\pm0.02299$ (6) & $0.00082\pm0.02088$ (4) & $0.00044\pm0.02295$ (5) \\
Smooth BH & $S\to G$ & 5:5 & $-0.02537\pm0.01602$ (9) & $-0.01076\pm0.01356$ (8) & $-0.01327\pm0.01485$ (8) \\
\addlinespace
Noisy BH & $G\to S$ & 1:9 & $0.04266\pm0.05436$ (2) & $0.00575\pm0.01111$ (4) & $0.00565\pm0.01120$ (4) \\
Noisy BH & $G\to S$ & 3:7 & $0.00216\pm0.06842$ (5) & $-0.00607\pm0.01411$ (8) & $-0.00958\pm0.01484$ (8) \\
Noisy BH & $G\to S$ & 5:5 & $-0.00375\pm0.08704$ (5) & $-0.00288\pm0.01178$ (5) & $-0.00436\pm0.01373$ (7) \\
Noisy BH & $S\to G$ & 1:9 & $0.00607\pm0.05314$ (5) & $-0.00494\pm0.01566$ (6) & $-0.00958\pm0.01709$ (6) \\
Noisy BH & $S\to G$ & 3:7 & $-0.06486\pm0.06476$ (9) & $-0.00649\pm0.00907$ (7) & $-0.01110\pm0.01046$ (9) \\
Noisy BH & $S\to G$ & 5:5 & $-0.03394\pm0.08360$ (8) & $-0.00009\pm0.01757$ (4) & $-0.00642\pm0.02093$ (6) \\
\bottomrule
\end{tabular}
\end{table}

The paired controls in Section~\ref{sec:offline_diagnostics} draw new prior-only training banks independently of the sequential banks. Direct/synthetic bank sizes are 1,000/9,000, 3,000/7,000 or 5,000/5,000, according to the source ratio. Both arms fit normalisation on the shared directly-simulated subset and use Adam with batch size 200 and learning rate $5\times10^{-4}$; no sequential-proposal correction is needed. Direct exposure is 1 in direct-only batches and $\rho=n_S/(n_S+n_G)$ in augmented batches.

Each arm retains 1,000 posterior draws restricted to prior support using a finite rejection budget. Diagnostic posterior MMD$^2$ uses the kernel above with coordinate scales and bandwidth estimated from the task's finite reference posterior. The mean marginal Wasserstein distance first divides each parameter coordinate by its prior width. Reference-mean error is the mean absolute coordinate difference between posterior and reference means. These definitions differ from the joint Wasserstein, legacy kernel and generating-parameter MAE used in Section~\ref{sec:metrics}; their numerical values must not be merged with the earlier workflow tables.

The full diagnostic data collection used 15,512 extra direct calls per task: 6,000 site trajectories, 9,000 prior-bank trajectories and 512 additional validation trajectories not used in the surrogate and augmentation checks reported here. The total 31,024 calls, diagnostic training and sampling costs are excluded from workflow speed-ups. Source records, bank identities, timings, acceptance rates and 420 posterior sample sets are retained in the evidence archive; weights and full simulation banks remain in the original experiment directories.

\FloatBarrier
\end{document}